\documentclass[12pt, titlepage]{article}
\newtheorem{theorem}{Theorem}[section]
\newtheorem{definition}{Definition}[section]
\newtheorem{lemma}{Lemma}[section]

\newtheorem{corollary}[theorem]{Corollary}
\newcommand {\dn}[1] {\boldsymbol #1}

\newcommand{\qed}{\nobreak \ifvmode \relax \else
      \ifdim\lastskip<1.5em \hskip-\lastskip
      \hskip1.5em plus0em minus0.5em \fi \nobreak
      \vrule height0.75em width0.5em depth0.25em\fi}

\usepackage{multirow}
\usepackage{psfrag}
\usepackage{epsf,epsfig}
\usepackage{latexsym,rotating,amsfonts,amsmath,amssymb,float}
\usepackage{subfigure}
\usepackage{psfrag,fancybox}
\usepackage[dvips]{color}
\usepackage{graphicx}
\usepackage{epstopdf}
\usepackage{amsmath}
\usepackage{epsfig}
\usepackage[sort&compress]{natbib}
\usepackage{natbib}
\usepackage{float}

\newcommand{\pref}[1]{%
    \ref{#1} \ifnum\count0=\pageref{#1}\relax%
    \else (page \pageref{#1})\fi}
\newcommand{\eref}[1]{%
        \ref{#1}\ifnum\count0=\pageref{#1}\relax%
        \else {, p.\pageref{#1}}\fi}
\newcommand{\comment}[1]{}
\newlength{\labwidth}

\newcommand{\logit}{\mbox{logit}}

\newcommand{\noi}{\noindent }

\begin{document}

\title{Explaining the behavior of joint and marginal Monte Carlo estimators
in latent variable models with independence assumptions.}
\author{Silia Vitoratou, Ioannis Ntzoufras and Irini Moustaki.}
%\date{}
% \rightheader{Monte Carlo error}\maketitle
\maketitle

 \abstract{

In latent variable models the parameter estimation can be implemented by using the joint or
the marginal likelihood, based on independence or conditional independence assumptions.
The same dilemma occurs within the Bayesian framework
with respect to the estimation of the Bayesian marginal (or integrated) likelihood,
which is the main tool for model comparison and averaging.
In most cases, the Bayesian marginal likelihood is a high dimensional integral that
cannot be computed analytically and a plethora of methods based on
Monte Carlo integration (MCI) are used for its estimation.
In this work, it is shown that the joint MCI approach
makes subtle use of the properties of the adopted model,
leading to increased error and bias in finite settings.
The sources and the components of the error associated with estimators under the two approaches are
identified here and provided in exact forms. Additionally,  the effect of the sample covariation
on the Monte Carlo estimators is examined. In particular, even under independence assumptions the sample covariance will be close to
(but not exactly) zero which surprisingly has a severe effect on the estimated values and their
variability. To address this problem, an index of the sample's divergence from independence
is introduced as a multivariate extension of covariance. The implications addressed here are important
in the majority of practical problems appearing in
Bayesian inference of multi-parameter models with analogous structures.
}

\newpage
\section{\textbf{Introduction}}
\label{sec:introMCE}

Latent variable models are widely used to capture latent constructs
by means of multiple observed indicators (items). From the early readings,
the methods applied for the parameter estimation of model settings
with latent variables relied either on the joint
(\citealp{LorNov68}; \citealp{Lord1980})
or the marginal likelihood (\citealp{BockLie70}; \citealp{BockAit81}). The former suggests to estimate
the observed and latent variable scores simultaneously while the latter
to marginalize out the latent variables prior to the model parameter estimation. Similarly,
counterpart approaches have been developed within the Bayesian context
(for instance  \citealp{Misl86}; \citealp{GiffSwa90}; \citealp{Kimetal94}; \citealp{Baker98}; \citealp{PatzJunk99}).

The Bayes factors, posterior model probabilities and the corresponding odds
\citep{kas:raf95} require the computation of the Bayesian marginal (or integrated) likelihood
which is defined as the expectation of the likelihood over the prior distribution.
To separate from the marginal likelihood term used in the context of latent variables, we will
refer to this as the Bayesian marginal likelihood (BML). In most cases the BML
 is a high dimensional integral which is not
analytically tractable. Sophisticated Monte Carlo techniques have been developed
throughout the years, such as the bridge sampling \citep{me:wo96} and the Laplace-Metropolis estimator
\citep{le:raf97}, among others. Despite of the method implemented however,
the BML can be estimated by considering either the joint
or the marginal likelihood expressions.

Intuitively, one expects the joint approach to be less efficient especially as the number of dimensions increases.
In this work, obtain analytical expressions for the variances associated with the estimator of each approach and
we consider the factors that influence their associated Monte Carlo error (MCE).
In particular, we illustrate graphically and mathematically that even though the MCE
is not by definition associated directly with the dimensionality
of a model, the latter plays a key role through the variance components.
In turn, the variance components are directly influenced by the number of the variables
involved and their variability. Additionally, we demonstrate  the effect of the sample
covariation on the Monte Carlo estimates, which is considerably understated in the literature.
In particular, for independent random variables the sample covariance is typically close but not exactly equal to zero.
Here, we illustrate that, in high dimensions, even small sample covariances
influence the estimators  producing biased Monte Carlo estimates.
This bias usually remains undetected, due to the fact that
the effect of sample covariation also causes underestimation of the corresponding MCEs.

Concerns arise with respect to convergence,
since the extensive use of simulation methods nowadays is not always followed
by the necessary precautions to ensure accurate estimation of the quantity of interest.
For instance, \cite{koeh09} reported that in a large number
of articles with simulation studies, only a tiny proportion provided
either a formal justification of the number of replications
implemented or the actual estimate of the Monte Carlo error (MCE).
That is, integral approximations are based on an arbitrary number of replications,
that are considered to be ``large enough"
to accurately estimate the quantity of interest. Nevertheless,
in complex high dimensional problems, where the rate of
convergence can be extremely low, millions of iterations may be required to
achieve a desirable level of precision for the MC estimate of interest.
Hence, in many cases the simulations are practically stopped ``when patience runs out", as \cite{Joetal06} fluently describe.
The remarks that are made in this paper facilitate the understanding of the error and
bias mechanism of Monte Carlo methods under independence and conditional independence and
hopefully will assist the researchers to accurately estimate the quantity of interest in high dimensions.

The structure of the paper is as follows. Section \ref{sec_gllvm} presents a
motivating example with regard to the estimation of the BML in a model with latent variables.
Three popular Markov Chain Monte Carlo (MCMC) methods are implemented, under both joint and marginal approaches.
Key observations are made based on the comparison of the derived estimated values which motivate further research.
Section \ref{sec:2estim} presents the Monte Carlo integration under the joint and marginal settings,
with emphasis on high dimensional integrals where independence can be assumed for the integrand.
The MCEs under both approaches are derived in Section \ref{subsec:2MCEs}
while the factors that affect the error
are considered in Section \ref{subsec:factorsMCEs}.
For illustration purposes a simple example is provided, that is,
estimating the mean of the product of
independent and identically distributed (i.i.d) $Beta$ random variables.
In Section \ref{subsec:Cond_MCEs}, the variance reduction in the case of conditional independence is discussed.
In Section \ref{sec:OnCov} the total covariation of $N$ variables is defined as a
multivariate counterpart of covariance. A corresponding index
that measures the sample's divergence from independence
is developed and employed to amplify the factors that influence the total sample covariation.
Finally, it is shown that in finite settings where the sample covariation is non zero,
the MCE associated with the joint approach is underestimated.

\section{\textbf{A motivating example: BML estimation in generalised linear latent trait models}}
\label{sec_gllvm}

A broad and popular family of models that can handle continuous, discrete and categorical
observed variables are the generalised linear latent variable models (GLLVM, \citealp{bart2011}).
Due to GLLVM's versatile applicability, they are utilized in this section
to amplify the difference between the joint and marginal likelihood approaches.
In particular, we focus on a latent trait model \citep{mou:kn00} with binary observed items,
under the Bayesian paradigm. The BML is computed in a simulated data set under the joint and marginal
approaches. The derived estimations raise specific concerns which are discussed at the end of the section.

\subsection{\textbf{Model setting and estimation techniques}}
\label{subsec:model}
The GLLVM consist of four main components:
(a) the multivariate random component  \linebreak
$\mathbf{Y}_i = (Y_{i1}, Y_{i2}, \dots, Y_{ip})$ of the response variables of subject $i$,
(b) a set of $k$ latent variables $\dn{Z}_i = (Z_{i1}, \dotsm Z_{ip})$ characterizing subject $i$,
(c) the linear predictor $\dn{\eta} _i = (\eta_{i1}, \dots, \eta_{ip})$ of the latent variables $\dn{Z}_i$ for subject $i$ and
(d) the link function $\upsilon (\cdot)$, that connects the previous three components.
Hence, a GLLVM can be summarized as
\begin{equation}
\label{glvm}
Y_{ij}\, \vert {\rm {\bf Z}}_i
\sim ExpF,
~~
\dn{\eta} _{i} = \dn{\alpha} + \dn{Z}_i \dn{\beta}^T,
~~ \dn{Z}_i \sim \pi(\dn{Z}_i)
\mbox{~~and~~}
\upsilon \big[  E(Y_{i }\vert {\rm {\bf Z}}_i\,) \big] = \dn{\eta}_i,
\mbox{~~for~~$i=1,\ldots N$},\nonumber
\end{equation}
where $ExpF$ is a member of the exponential family,
are the $k$ latent variables for the $i$ subject,
$\dn{\alpha}=(\alpha_1, \dots, \alpha_p)$,
$\dn{\beta}=( \beta_{j\ell} \, ; j=1,\dots, p, \ell=1,\dots, k)$
and $\dn{Z}_i \sim \pi(\dn{Z}_i)$ denotes that $\dn{Z}_i$ is random variable with density $\pi(\dn{Z}_i)$.
In the above formulation, $\pi(\textbf{Z}_i)$ needs to be specified for the latent variables.
Typically, the latent variables are assumed to be
a-priori distributed as independent standard normal
distributions, that is, $\dn{Z}_i \sim N( \dn{0}, \dn{I}_p )$ for all individuals \citep{bart2011},
where $\dn{I}_p$ is the identity matrix of dimension $p \times p$.

In the following, we focus on models with binary responses and $k$ latent variables, which belong to the family of
generalized latent trait models discussed in \cite{mou:kn00}. The logistic model is used for the response
probabilities:
\begin{eqnarray}
\logit \big[ P_{ij}\,(\dn{Z}_i) \big]=\alpha_j+\overset{k}{\underset{\ell~=1}{\sum}}\beta_{j\ell }\,Z_{i\ell},
 ~ i=1,\ldots,N,~j=1,\ldots p,\nonumber
\end{eqnarray}
\noi where $P_{ij}(\dn{Z}_i)$ is the conditional probability of a
positive response by the individual $i$ to item $j$.
The model assumes that the responses are independent given the latent variables $\dn{Z}=( Z_{ij}; \, i=1,\dots,N, \, j=1,\dots,p )$
(\emph{local independence} assumption) leading to either the joint  (\citealp{LorNov68}; \citealp{Lord1980})
\begin{eqnarray}
\label{lik.joint}
f(\textbf{Y}|\,\boldsymbol{\theta},\textbf{Z})= \prod_{i=1}^N f(Y_i|\,\boldsymbol{\theta},\dn{Z}_i)= \prod_{i=1}^N \prod_{j=1}^p P_{ij}\,(\dn{Z}_i)^{y_{ij}}\, \big[ 1-P_{ij}\,(\dn{Z}_i) \big]^{(1-y_{ij})}
\end{eqnarray}
\noi or the marginal likelihood (\citealp{BockLie70}; \citealp{BockAit81}; \citealp{mou:kn00})
\begin{eqnarray}
\label{lik.marg}
f(\textbf{Y}|\,\boldsymbol{\theta} )= \prod_{i=1}^N \int \prod_{j=1}^p P_{ij}\,(\dn{Z}_i)^{y_{ij}}\,[1-P_{ij}\,(\dn{Z}_i)]^{(1-y_{ij})}\,d\,\dn{Z}_i,
\end{eqnarray}
\noi where $\dn{\theta}=( \dn{\alpha}, \dn{\beta} )$.

For the Bayesian counterpart of the model, priors distributions are additionally assigned on model parameters $\dn{\theta}$.
The prior specification of the model used here is based on the ideas presented
by \cite{ntzetal:03} and further explored in the context of
generalized linear models by \citet[equation 6]{fousk:09}. For a
GLLVM with binary responses, this prior corresponds to a
$N(0,4)$. In the case of $k>1$ latent variables, constraints need to be imposed on the loadings
$\boldsymbol{\beta}$ to ensure identification of the model. To achieve a unique solution, the loadings matrix
is constrained to be a full rank lower triangular matrix
(see also \citealp{geZou:96}, \citealp{agwest00} and \citealp{lowest:04}), by setting
$\beta_{j\ell }=0$ for all $j < \ell$ and $\beta_{jj}>0$. The prior is summarized as follows:

$$
\pi(\beta_{j\ell}) = \left\{
\begin{array}{lll}
LN(0,1) & \mbox{if} & j=\ell \\
N(0,4) & \mbox{if} & j>\ell
\end{array}
\right.
$$

\noi where $X\sim LN(0, 1)$ is the log-normal distribution with zero mean and the variance equal to one for
$\log\!X$.
For diagonal elements $\beta_{jj}$, the $LN(0,1)$ was selected as a prior in order to approximately match the prior standard deviation
used for the rest of the parameters.
Moreover, this is one of the default prior choices for such parameters in the relevant literature;
see for example in \cite{KanCo07} and references therein.

In analogy with (\ref{lik.joint}) and (\ref{lik.marg}), under the local independence assumption there are two equivalent
formulations of the BML, namely
 \begin{eqnarray}\label{integrlik.joint}
\displaystyle f(\textbf{Y}) = \int \prod_{i=1}^{N}f(Y_{i}|\,\boldsymbol{\theta},\dn{Z}_i)  \, \pi( \dn{\theta}, \textbf{Z} ) \, d (\dn{\theta}, \textbf{Z})
 \end{eqnarray}
and
 \begin{eqnarray}\label{integrlik.marg}
\displaystyle f(\textbf{Y})
= \int f( \dn{Y} | \,\dn{\theta}) \, \pi( \dn{\theta}) \,d\dn{\theta}
= \int \left[\prod_{i=1}^{N} \int f(Y_i|\,\boldsymbol{\theta},\dn{Z}_i) \, \pi(  \dn{Z}_i) \, d\dn{Z}_i \right]  \,
       \pi( \dn{\theta} ) \, d\dn{\theta} ~.
 \end{eqnarray}

\noi Hereafter we  refer to (\ref{integrlik.joint}) with the term joint approach  and to (\ref{integrlik.marg}) with the term marginal approach  for
the BML and we compare them within the Bayesian framework.

For both approaches, we employ  three popular BML estimators  namely:
the reciprocal mean estimator ($RM$; \citealt{GelDey94}),
the  bridge harmonic estimator ($BH$; \citealt{me:wo96}, often refer to as the generalised harmonic mean)
and the  bridge geometric estimator ($BG$; \citealt{me:wo96}).
The identities that correspond to these estimators are provided in the Appendix. In
order to construct the estimators using the joint approach, the parameter vector is
augmented to include the latent variables, that is \linebreak
$\boldsymbol{\vartheta}=\{\boldsymbol{\theta}, \bf{Z}\}=\{\boldsymbol{\alpha}, \boldsymbol{\beta}, \bf{Z}\}$,
while for the marginal approach it holds $\boldsymbol{\vartheta}=\boldsymbol{\theta}=\{\boldsymbol{\alpha}, \boldsymbol{\beta}\}$.

The estimators require also an \emph{importance} function
$g(\boldsymbol{\vartheta})$. The objective and recommendation of many authors
\citep{me:wo96,diccetal97,ge:me98,me:sch02}, is to choose a
density similar to the target distribution (here the posterior).
In the current example,
we use an approximation based on the posterior moments for each parameter, with structure
$ g(\boldsymbol{\theta})=g(\boldsymbol{\alpha})g(\boldsymbol{\beta}_{e})$
where
$$
g(\boldsymbol{\alpha}) \sim MN(\widetilde{\textbf{m}}_{\boldsymbol{\alpha}},\widetilde{\boldsymbol{\Sigma}}_{\boldsymbol{\alpha}}) \mbox{~~and~~}
g(\boldsymbol{\beta}_{e})\sim MN(\widetilde{\textbf{m}}_{\boldsymbol{\beta}_e},\widetilde{\boldsymbol{\Sigma}}_{\beta_e}),\, \dn{\beta}_{e}=\beta_{j\ell},~ {i \geq \ell}
$$
and $\boldsymbol{\beta}_{e}$
refers to the non-zero components of $\boldsymbol{\beta}$ with elements
$\log\beta_{jj}$  for  $j=1,\dots,p$ and $\beta_{j\ell}$ for~ $j>\ell$.
The $MN(\widetilde{\textbf{m}}, \widetilde{\boldsymbol{\Sigma}})$ denotes a multivariate normal distribution
whose parameters ($\widetilde{\textbf{m}}, \widetilde{\boldsymbol{\Sigma}}$)
are the posterior mean and variance-covariance matrix estimated from the MCMC output.
For the joint approach,  the $ g(\boldsymbol{\vartheta})$ is simply augmented for the latent vector

$$
g(\boldsymbol{\vartheta})=g(\boldsymbol{\alpha})g(\boldsymbol{\beta}_e)\prod_{\ell=1}^k\prod_{i=1}^N g(Z_{i\ell}),
$$

\noi where $g(Z_{i\ell}) \sim N( \tilde{m}_{Z_{i\ell}}, \tilde{s}^2_{Z_{i\ell}})$,
with parameters estimated from the MCMC output used to approximate the posterior $\pi( Z_{i \ell}|\dn{Y})$.

\subsection{\textbf{Simulation study}} \label{subsec:simGLLVM}

A simulated data set with $p=6$ items, $N=600$ cases and $k=2$ factors is firstly considered.
The model parameters were selected randomly from a uniform distribution U(-2,2).
Using a Metropolis within Gibbs algorithm, 50,000 posterior observations were
obtained after discarding a period of 10,000 iterations and considering
a thinning interval of 10 iterations to diminish autocorrelations. The posterior
moments involved in the construction of the importance function
were estimated from the final output and an additional sample of equal size was
generated from  $g(\dn{\theta})$. The MCMC estimators  were computed in two versions, joint and marginal,
using the entire MCMC output of 50,000 iterations. In a second step, the simulated sample
was divided into 50 batches (of 1,000 iterations) and
the integrated log-likelihood was estimated at each batch.
The standard deviation of the log-BML estimators over the different
batches is considered here as its MCE estimate (\citealt{Schm:82}, \citealt{bratetal:1987}, \citealt{CarLouis00}).

In this example, the BML (\ref{integrlik.marg}) was calculated
by approximating the inner integrals  with fixed Gauss-Hermite quadrature points. This way,
the computational burden is considerably reduced without compromising
the accuracy, since such approximations are fairly precise in low
dimensions. Other approximations can be alternatively used, such
as the adaptive quadrature points
(\citealt{rabe.skrondal.pickles:05}, \citealt{schilling.Bock:05})
or Laplace approximations \citep{huber.ea:04}.
All simulations were conducted using $\mathrm{R}$ (version 2.12) on a quad core i5
Central Processor Unit (CPU), at 3.2GHz and with 4GB of RAM.
The estimated values for each case are presented in Table \ref{table:GLLVM_margjoint}.

\begin{table}[h!]
\caption{\small \textbf{BML estimates (log scale) for the GLLVM example}}
\label{table:GLLVM_margjoint}
\begin{center}
\begin{tabular} {ccccc}
Approach & Estimator & Estimation   & Batch mean  &  $M\widehat{C}E$   \\
\hline
\hline
\\
        & RM & -2062.3  & -2053.9 & ~~3.46  \\
Joint   & BH   & -2068.8  & -2065.5 &  17.92   \\
        & BG  & -2073.3  & -2072.8 & ~~2.21   \\
\\
\hline
\\
        & RM & -2071.3  & -2071.2 &  0.28   \\
Marginal& BH   & -2069.6  & -2069.3 &  2.11   \\
        & BG  & -2071.6  & -2071.6 &  0.07   \\
\hline
\multicolumn{5}{p{10.5cm}}{\scriptsize
        The estimated BML of a GLLVM model with $p=6$ items, $N=600$ cases and $k=2$ factors. Each estimation
        was computed over a sample of 50,000 simulated points while the batch mean  and the associated error were computed over 50 batches of
        1,000 points each. RM: Reciprocal mean estimator, BH: Bridge harmonic estimator and BG: Bridge geometric estimator.       }

\end{tabular}
\end{center}
\end{table}

\subsection{\textbf{Estimations and key observations}}\label{subsec:keyObs}

\begin{figure}[hp]
  \begin{center}\vspace{1cm}
   {\subfigure[$BH_{_J}$  and $BG_{_J}$  ]{\label{fig:bridge_a}
                                 \psfrag{m}[c][c][0.6]{\sf BML estimators (log scale)}
                                 \includegraphics[width=0.32\textwidth ]{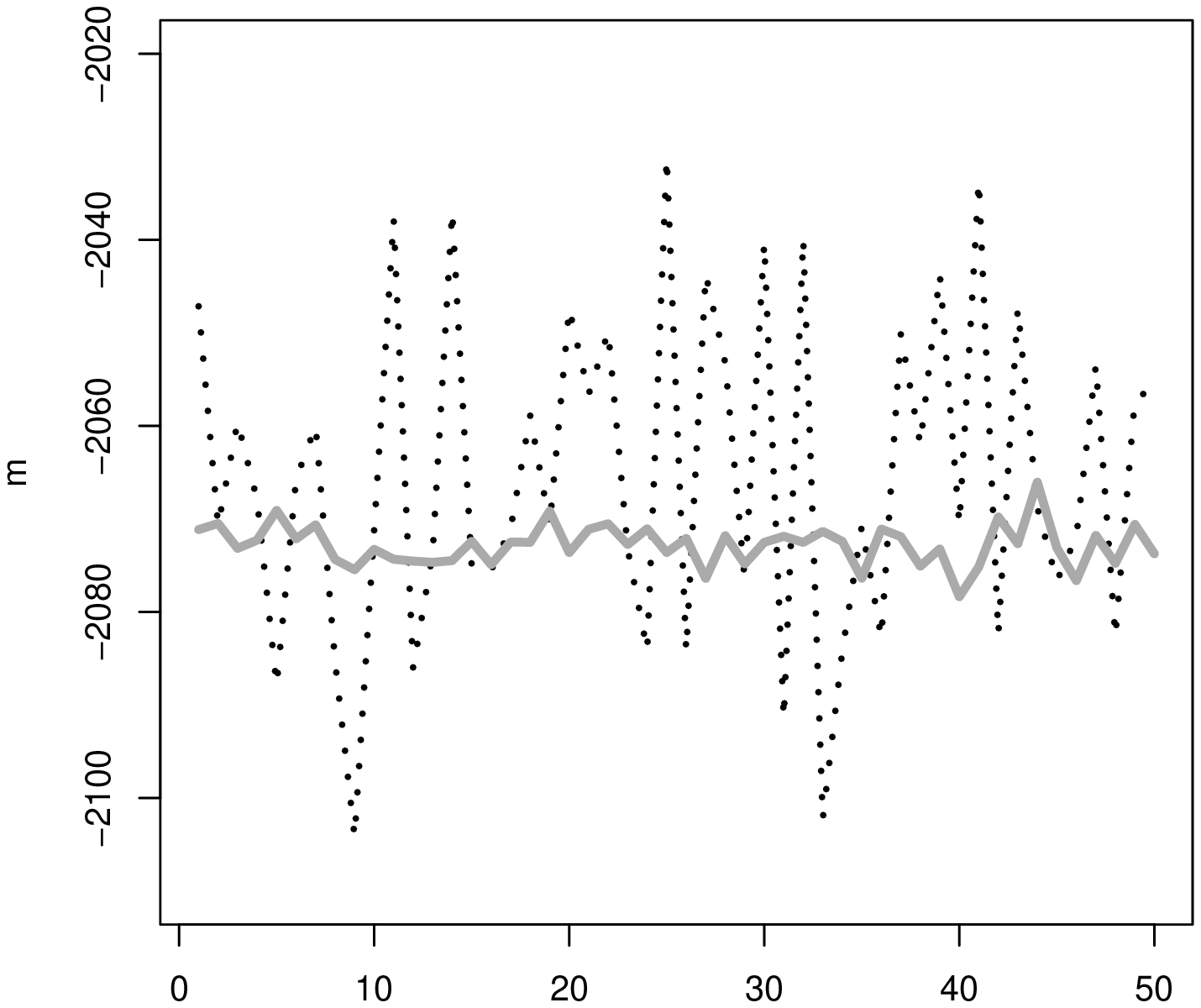}}}
   {\subfigure[$BG_{_J}$  and $BG_{_M}$  ]{\label{fig:bridge_b}
                               \psfrag{a}[c][c][0.9]{\sf Generalized Harmonic and Geometric bridge sampling estimators}
                                \psfrag{m}[c][c][0.8]{\sf Batch (size = 1000 per batch)}
                                \includegraphics[width=0.32\textwidth ]{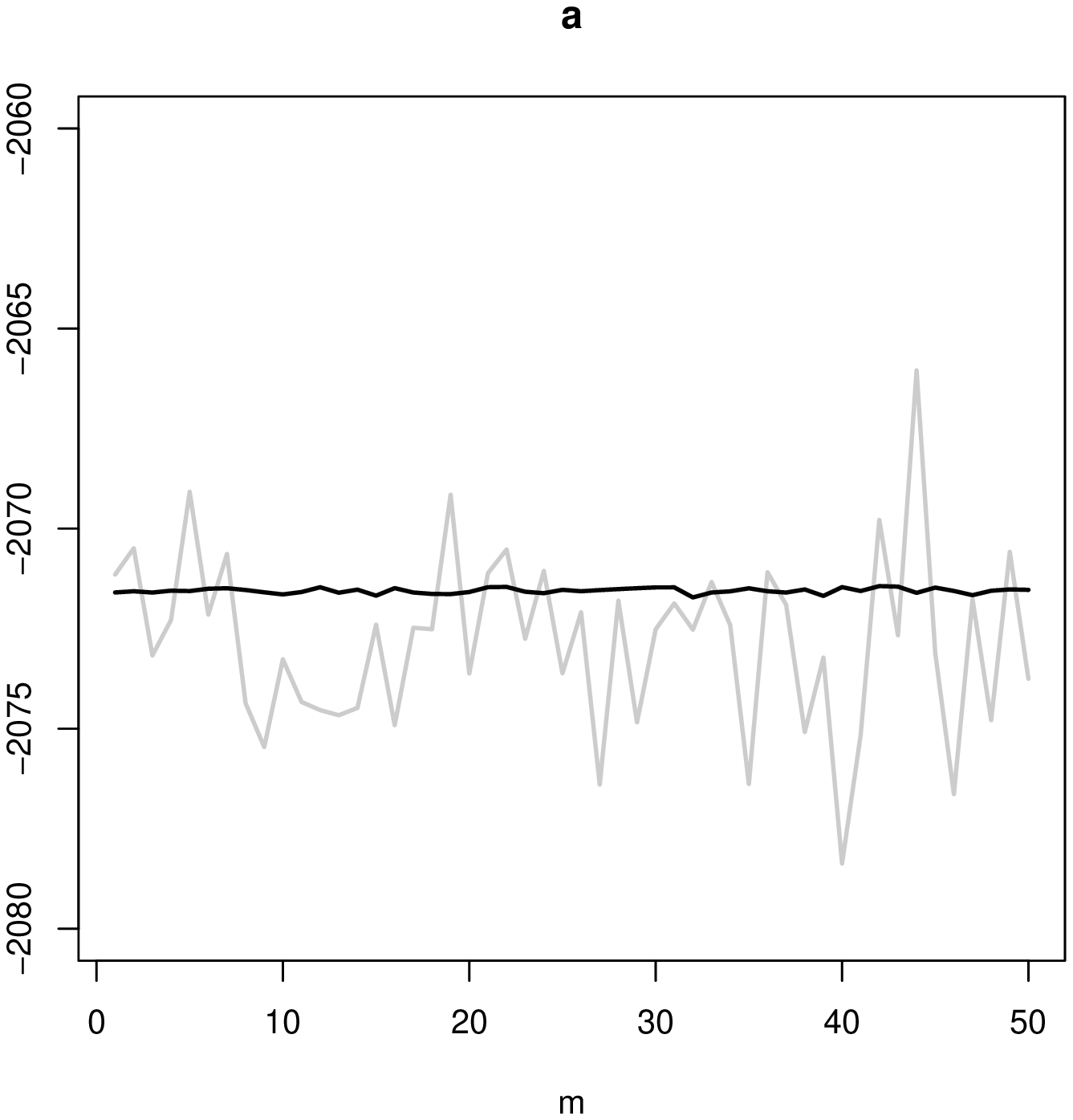}}}
  {\subfigure[$BH_{_J}$, $BG_{_J}$  and $BG_{_M})$]{\label{fig:bridge_c}
                                 \includegraphics[width=0.32\textwidth ]{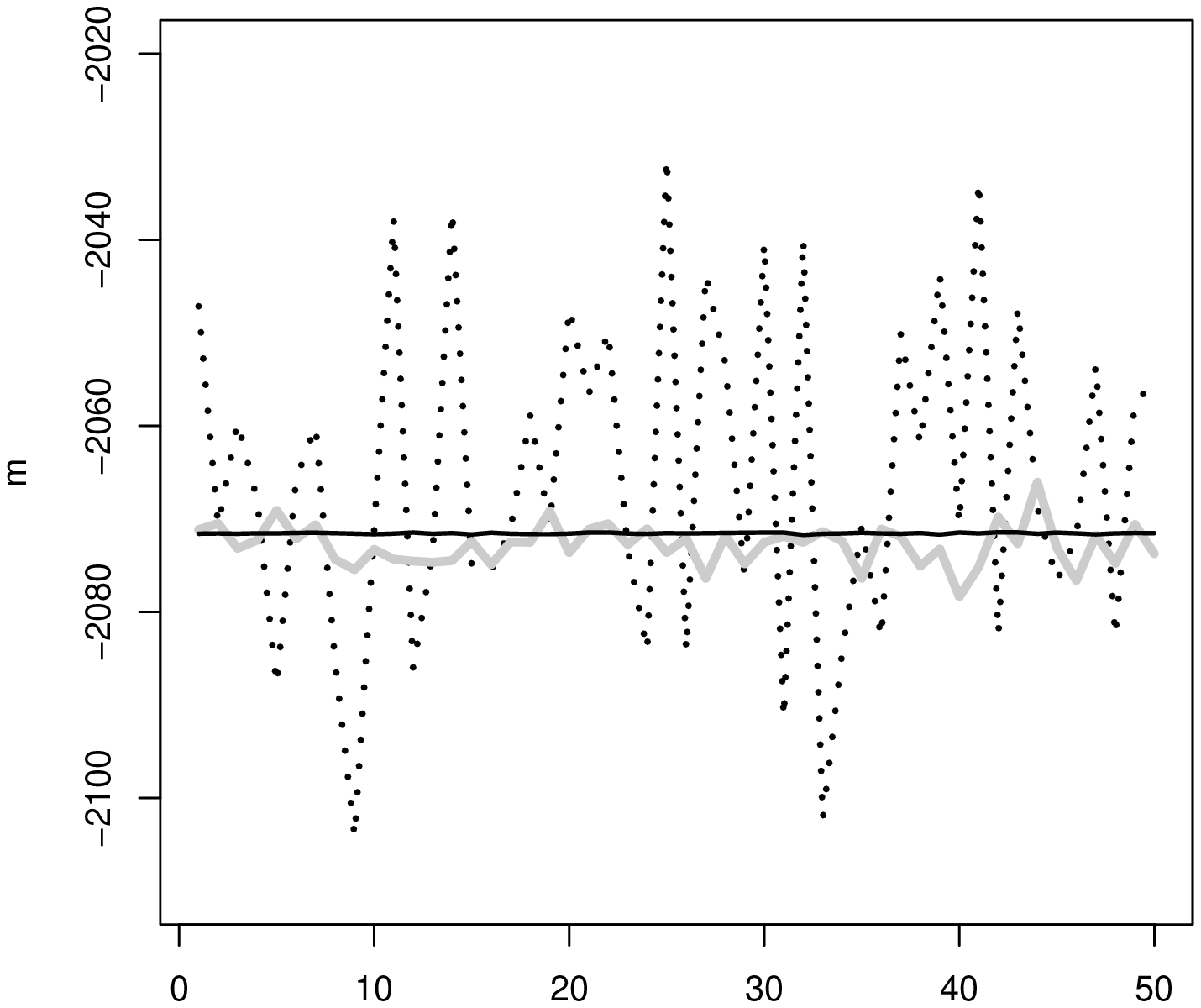}}}
    \end{center}
        \caption{\scriptsize The \emph{joint} bridge harmonic estimator $BH_{_J}$  (dotted line),
                             the \emph{joint} bridge geometric  estimator $BG_{_J}$ (gray solid line)
                             and the \emph{marginal} bridge geometric  estimator $BG_{_M}$  (black solid line),
                             for the BML (log scale), implementing a simulated data set with
                             $p=6$ binary items, $N=600$ cases and $k=2$ factors, over 50 batches.}
\end{figure}

The first observation derived from the current example refers to the variability differences between the estimators and between their
joint and marginal counterparts. For illustration purposes we focus on the two bridge sampling estimators.
The joint bridge harmonic $(BH_J)$ and bridge geometric $(BG_J)$ estimators are
depicted in Figure \ref{fig:bridge_a} over the 50 batches. The variability differences between them
is striking, implying that the geometric estimator is a variance reduction technique
as opposed to the  harmonic. The next step in our investigation was to compare the less variant
estimator with its marginal counterpart. Figure  \ref{fig:bridge_b} illustrates
that further variance reduction can be achieved by implementing the marginal
rather than the joint geometric estimator. It becomes apparent that even the efficient bridge geometric estimator
was considerably improved by employing the marginal approach. That fact is typical in high dimensional
models and often expected intuitively.

\begin{figure}[hp]
  \begin{center}\vspace{1cm}
   {\subfigure[Reciprocal estimator]{\label{fig:gllvmRec}
                                 \psfrag{a}[c][c][0.6]{\sf BML estimators (log scale)}
                                    \psfrag{b}[c][c][0.8]{\sf }
                                 \includegraphics[width=0.32\textwidth ]{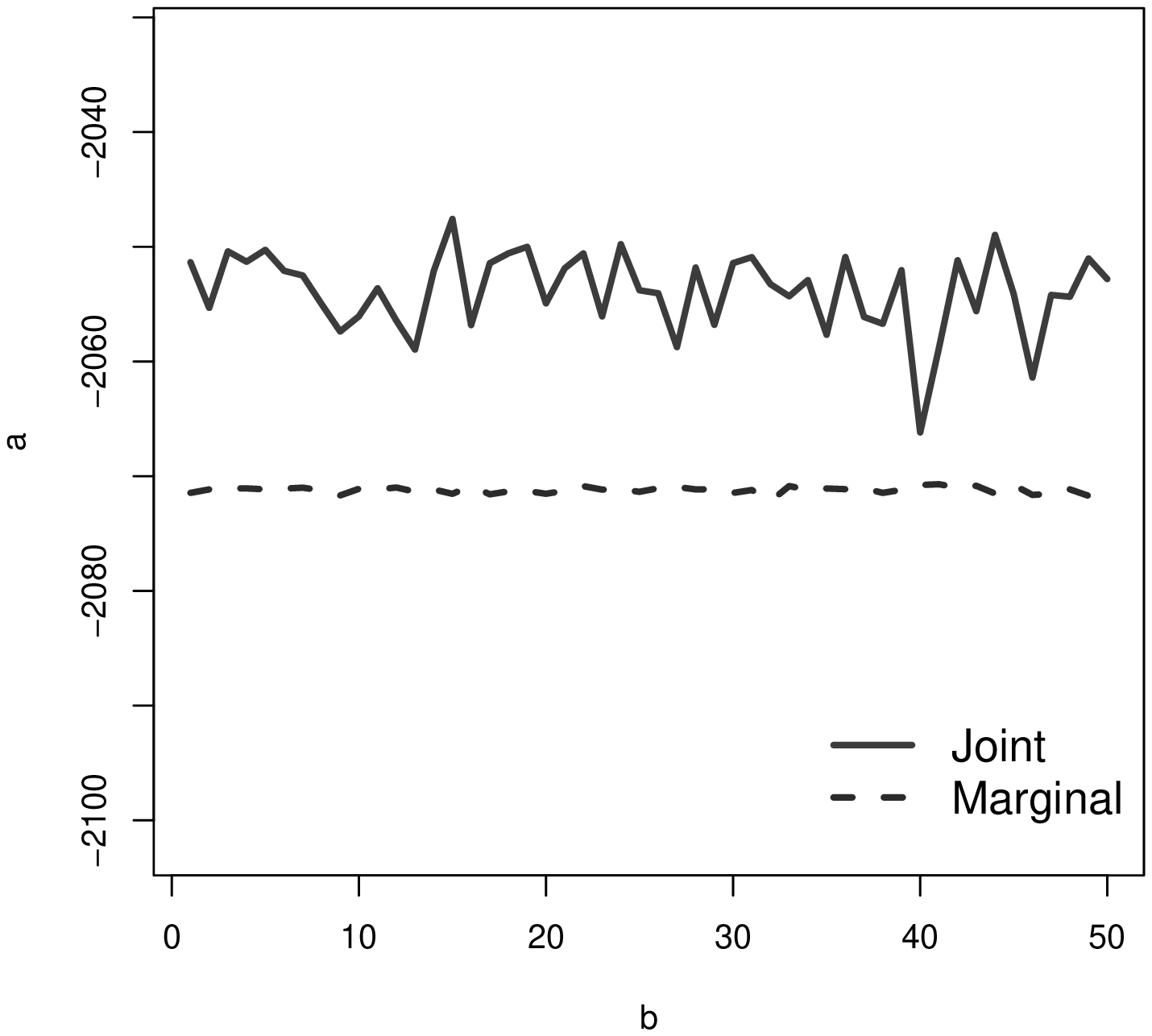}}}
   {\subfigure[Marginal estimators]{\label{fig:gllvmJoints}
                                \psfrag{b}[c][c][0.8]{\sf Batches ({batch size = 1000})}
                                \includegraphics[width=0.32\textwidth ]{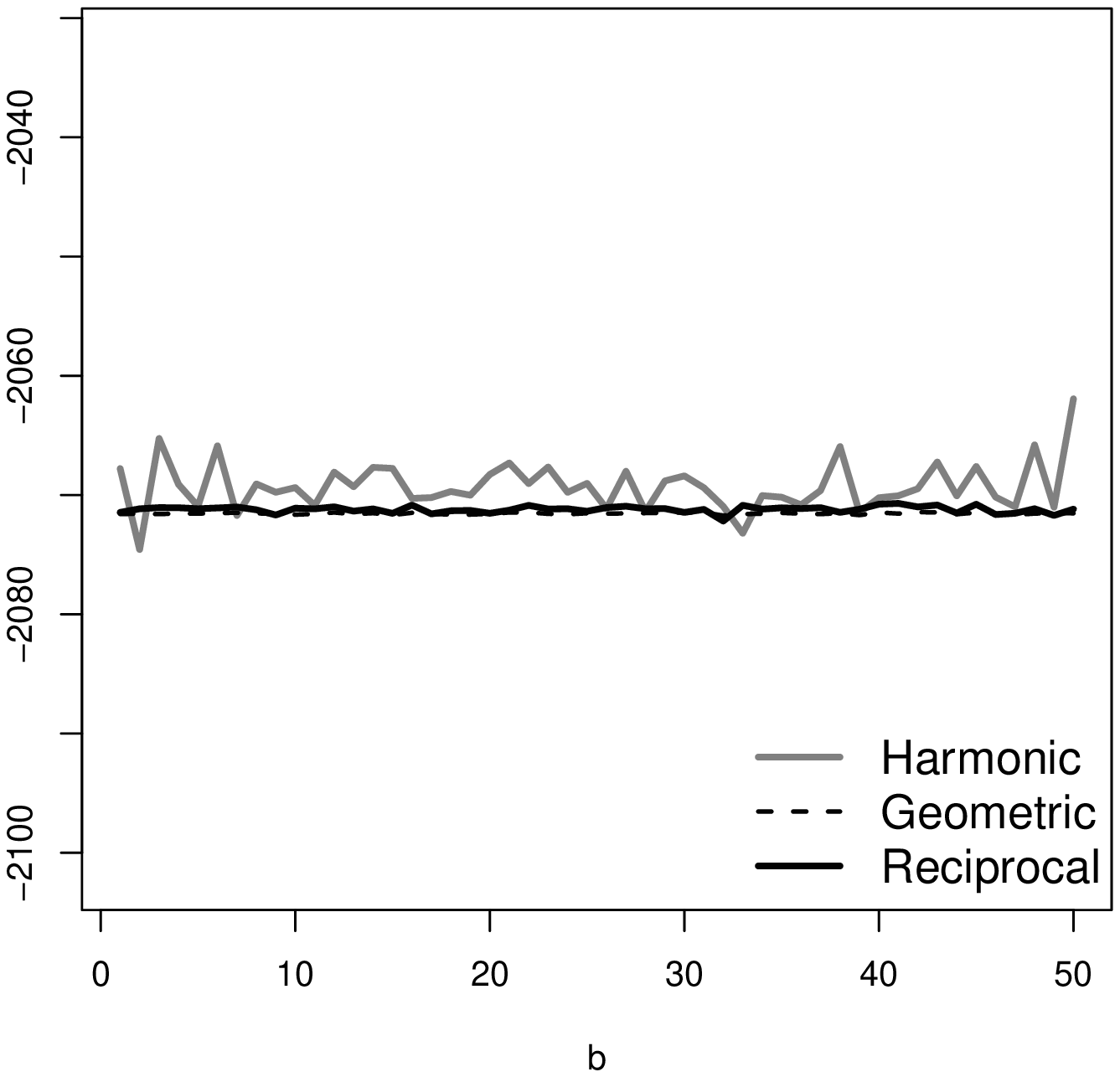}}}
  {\subfigure[Joint estimators]{\label{fig:gllvmMargs}
                                 \psfrag{b}[c][c][0.8]{\sf }
                                 \includegraphics[width=0.32\textwidth ]{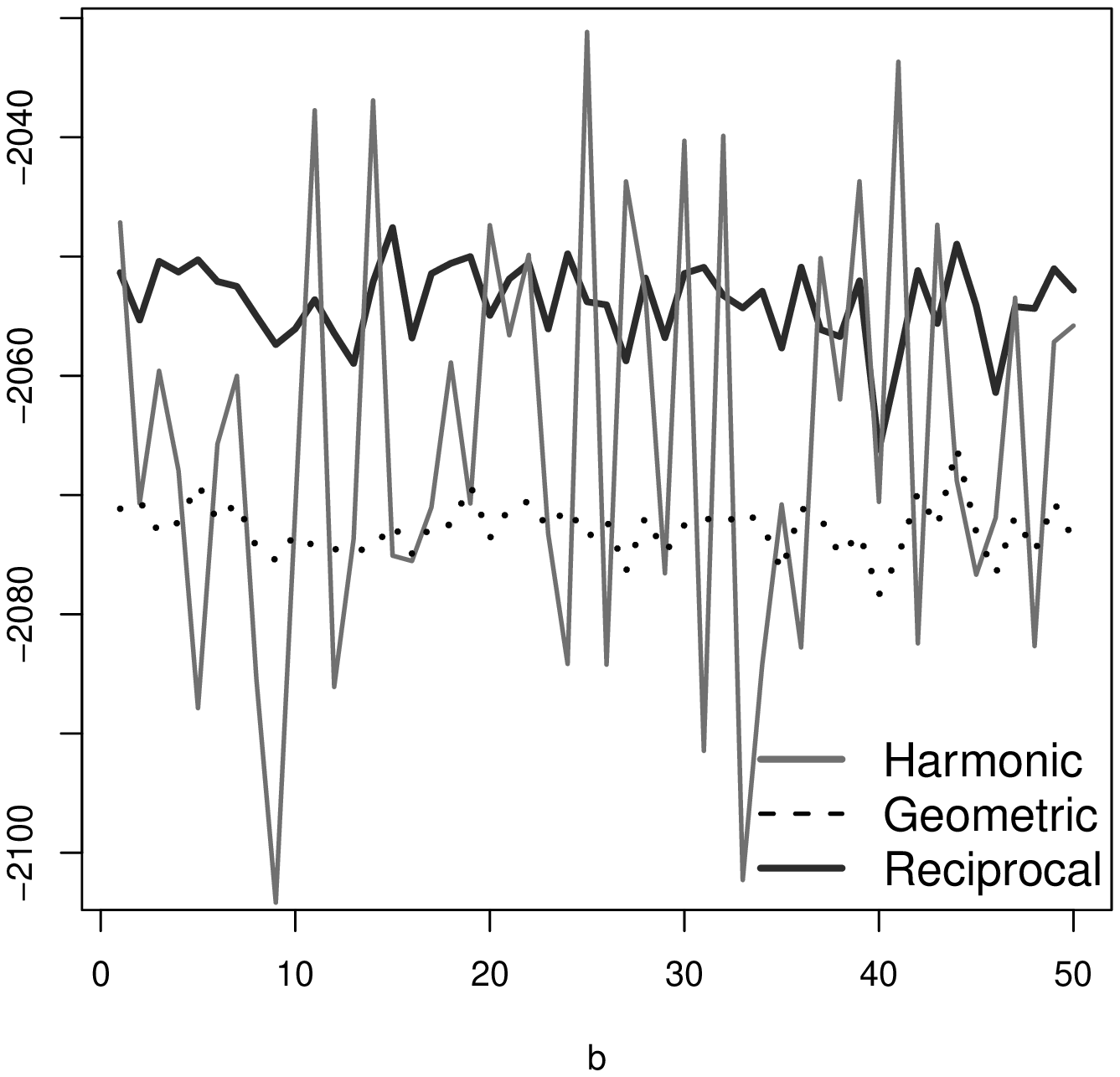}}}
    \end{center}
\caption{\scriptsize Joint and marginal approaches for the reciprocal ($RM$), bridge harmonic ($BH$)
and bridge geometric  ($BG$) estimators of the BML (log scale), implementing a simulated data set with
$p=6$ binary items, $N=600$ cases and $k=2$ factors, over 50 batches.}
\end{figure}

The second observation was less imaginable and it refers to the estimated values per se.
In particular, Figure  \ref{fig:bridge_c}
illustrates that the $BH_J$,  $BG_J$ and
$BG_M$  estimators vary  around a common estimated value for the BML and the divergencies present in Table \ref{table:GLLVM_margjoint}
are within the margins of their corresponding errors.
However this is not true in the case of the reciprocal estimator.
As opposed to the bridge  estimators, Figure \ref{fig:gllvmRec} illustrates that
substantially distant estimations were derived by the joint ($RM_J$)  and  marginal ($RM_M$)
reciprocal estimators. The difference in the estimated values
is about 10 units in log-scale, meaning that it far exceeds the corresponding MCEs and
hence cannot be explained  solely by variability. In addition, it is interesting to notice that the
$RM_J$ occurs to be much more divergent than the $BH_J$,
even though the latter is associated with 5 times higher error (Table \ref{table:GLLVM_margjoint}).
The three joint estimators are depicted in
Figure \ref{fig:gllvmJoints} and their marginal counterparts are illustrated in Figure \ref{fig:gllvmMargs}.

%\newpage

Several concerns arise therefore with regard to the convergence of the estimators in finite settings,
listed below:
\begin{itemize}
    \item [a)] What is the mechanism which produces these differences?
    \item [c)] Can the differences in the error be ameliorated to some extend by increasing the simulated sample size in finite settings?
    \item [d)] By increasing the number of the simulated points, do the discrepancies in the estimated values reduce? Where is this type of bias coming from?
\end{itemize}

Regarding the mechanism, we state here that is  related to the model assumptions.
Specifically, consider the model parameters
$\boldsymbol{\theta}$ fixed in the BML expressions (\ref{integrlik.joint}) and (\ref{integrlik.marg}).
It occurs that the joint expression implements the mean of the product
of the independent variables $f_{\,\boldsymbol{\theta}}(\textbf{Y}_i|\dn{Z}_i) $
while the marginal expression employs the product of their means.
The former is a generally applicable approach while the latter
occurs explicitly under independence. We conclude that the joint
approach makes subtle use of the local independence assumption.
This fact has direct implications on the estimated value and the associated error
which are thoroughly examined in the following section.

\section{\textbf{Joint and marginal Monte Carlo estimators under independence assumptions}} \label{sec:2estim}

The Monte Carlo integration techniques are reviewed here in a general framework,
since the subsequent theoretical findings extend beyond models with latent variables.
In particular, we consider any multi-dimensional integral of the form
\begin{equation}\label{MCmulti}
I=\int \phi(\textbf{Y})h(\textbf{Y})\,d\textbf{Y}, ~ \mbox{where}~ \textbf{Y}=(Y_1,Y_2,...,Y_N).
\end{equation}
\noi The MC approximation of the integral (\ref{MCmulti}) corresponds to the expected
value of $\phi(\dn{Y})$ over $h(\dn{Y})$.
Specifically, if $\dn{y}^{(r)}=  \big( y_1^{(r)},y_2^{(r)},...,y_N^{(r)}\big)$
and
$\textbf{\textbf{y}}^R=\big\{ \dn{y}^{(r)} \big\}_{r=1}^R$
is a random sample of points generated from the distribution $h( \dn{y} )$, then the estimator
$\widehat{I}=\overline{\phi}=\frac{1}{R} \sum_{r=1}^R \phi \big( \dn{y}^{(r)} \big)$
will approach (\ref{MCmulti}) for sufficiently large sample size $R$.
The degree of accuracy associated with the Monte Carlo estimator is directly related to the size of the simulated sample $R$.
The standard deviation  of $\overline{\phi}$  is the MCE of the estimator.
The MCE is therefore defined as the standard deviation of the estimator across simulations of the same number of replications $R$
and is given by:
\begin{equation}
\label{MCEdefin}
MCE = \sqrt{ \mbox{Var}(\,\overline{\phi}\,) } =\frac{\sigma_\phi}{ \sqrt{R} },\nonumber
\end{equation}
while an obvious estimator of MCE is given by
$\widehat{MCE}= S_\phi /\sqrt{R}$,
provided that an estimator $S_\phi^2$ of the integrand's variance $\sigma_\phi^2$ is available.
From (\ref{MCEdefin}), it occurs that the MCE directly depends on $\sigma_\phi$ and $R$.

Here we focus on the estimation of the expected value of
$\phi(\dn{Y})=\prod_{i=1}^N  \phi_i(Y_i)$ given by
\begin{equation}
\label{basicexpJoint}
I = E[\phi(\textbf{Y})]
  = E\left[ \prod_{i=1}^N  \phi_i(Y_i)\right ]
  = \int\prod_{i=1}^N  \phi_i(Y_i)h(Y_1,Y_2,...,Y_N)\,d(Y_1,Y_2,...,Y_N)~.
\end{equation}
Under
the assumption of independence for $Y_i$, we can rewrite (\ref{basicexpJoint}) as
\begin{equation}
\label{basicexpMarg}
I = \prod_{i=1}^N E\left[ \phi_i(Y_i)\right]
  = \prod_{i=1}^N \int\phi_i(Y_i)h_i(Y_i)dY_i~.
\end{equation}
The expressions (\ref{basicexpJoint}) and (\ref{basicexpMarg}) can be used to construct two
unbiased Monte Carlo estimators of $I$, described in Definitions \ref{def1} and \ref{def2} that follow.

\begin{definition}
\label{def1}
{\bf Joint estimator of $I$}. For any random sample $\big\{y_1^{(r)},y_2^{(r)},...,y_N^{(r)}\big\}_{r=1}^R$
 from $h$, the {\it joint estimator} of $I$ is defined as
\begin{equation}
\label{JointEst}
\widehat{I}_{J}= \overline{\phi} =  \frac{1}{R} \sum_{r=1}^R \phi \Big(y_1^{(r)},y_2^{(r)},...,y_N^{(r)}\Big)
               = \frac{1}{R} \sum_{r=1}^R \left[ \prod_{i=1}^N \phi_i \Big( y_i^{(r)}\Big) \right]~.
\end{equation}
\end{definition}

\begin{definition}
\label{def2}
{\bf Marginal estimator of $I$}. For any random sample $\big\{y_1^{(r)},y_2^{(r)},...,y_N^{(r)}\big\}_{r=1}^R$
 from $h$, the {\it marginal estimator} of $I$ is defined as
\begin{equation}
\label{MargEst}
\widehat{I}_{M} = \prod_{i=1}^N \left[ \frac{1}{R}\sum_{r=1}^R  \phi_i\Big( y_i^{(r)} \Big)\right]
                = \prod_{i=1}^N \overline{\phi}_i.
\end{equation}
\end{definition}
\bigskip

\noi
In the remaining of the paper we examine the divergencies between the two estimators
in finite settings, as a result of disregarding the  assumption of independence.

\subsection{\textbf{Monte Carlo errors}} \label{subsec:2MCEs}
\label{subsub_MCE}

The exact MCEs for the joint and marginal estimators
are expressed in terms of their variances. In particular, the variance
of the joint estimator (\ref{JointEst}) is directly linked to the
variance of the product of $N$ independent variables since
\begin{eqnarray}
Var(\widehat{I}_{J})
= Var\bigg[\frac{1}{R} \sum_{r=1}^R \Big\{ \prod_{i=1}^N \phi_i \big( y_i^{(r)}\big) \Big\} \bigg]
=\frac{Var\left[ \prod_{i=1}^N \phi_i(Y_i)\right]}{R}.
\label{var1}
\end{eqnarray}
On the other hand, the variance of the marginal estimator (\ref{MargEst}) is given by the
variance of the product of $N$ univariate MC estimators, that is
\begin{eqnarray}
Var(\widehat{I}_{M})=Var\left[\prod_{i=1}^N \overline{\phi}_i\right].
\label{var2}
\end{eqnarray}
The difference between (\ref{var1}) and (\ref{var2}) becomes apparent if
the early findings of \cite{good62} are reviewed within the framework of Monte Carlo integration.
\citet[eq. 1 and 2]{good62} provides  the variance $\sigma^2$ of the product of $N$
independent variables $Y_i$, $(i=1,...,N)$ with probability or density functions $h_i(Y_i)$.
For our purposes, we expand it to the case of functions $\phi_i(Y_i)$ of the original independent random variables,
leading to
\begin{eqnarray}
\label{variGOOD}
Var\left(\prod_{i=1}^N  \phi_i(Y_i)\right)
&=& \sum_{i=1}^{N} V_i\overset{N}{\underset{i\,'\neq i}{\prod}}E_{i\,'}^2+\overset{N}{\underset{i_1<\,i_2}{\sum}}V_{i_1}V_{i_2}\overset{N}{\underset{i\,'\neq i_1,\,i_2}{\prod}}E_{i\,'}^2 +...+V_1V_2 \cdot \cdot \cdot V_N,
\end{eqnarray}
\noi where $E_{i\,'}=E[\phi_{i\,'}(Y_{i\,'})]$ and $V_i=Var[\phi_i(Y_i)]$,
($i,i\,'\in \{1,...,N\}$), with all moments being calculated over the
corresponding densities $h_i(Y_i)$.

Equation (\ref{variGOOD}) can be written  as
\begin{eqnarray}\label{variGOOD2}
Var\left(\prod_{i=1}^N  \phi_i(Y_i)\right)
&=& \sum_{k=1}^{N} \sum_{ {\cal C} \in {{\cal N}\choose{k}} }
\left[ \prod_{i \in {\cal C} } V_i \prod_{j \in {\cal N} \setminus {\cal C} } E_j^2 \right],
\end{eqnarray}
where ${{\cal N}\choose{k}}$ is the set of all possible combinations of $k$ elements of ${\cal N}=\{1,2,\dots,N\}$
and any product over the empty set is specified to be equal to one.

The variances of the two Monte Carlo estimators in (\ref{var1})
and (\ref{var2}) may now be expressed in terms of (\ref{variGOOD}). Specifically,
the variance of the joint estimator is directly obtained by dividing the integrand's variance in
(\ref{variGOOD}) with the simulated sample size $R$.
For the marginal estimator, the variance  (\ref{var2}) can be obtained by substituting
$V_i$ by $V_i/R$ in (\ref{variGOOD2}).
The variance components that correspond to the MCEs in each case are presented in the following lemma.

\begin{lemma}
\label{prop:EstVars}
The  variances of the joint (\ref{JointEst}) and  marginal estimators (\ref{MargEst}) are given by
\begin{eqnarray}
\label{var1good}
Var(\widehat{I}_{J})
%&=& \frac{\overset{N}{\underset{i=1}{\sum}} V_i\overset{N}{\underset{i\,'\neq i}
%                     {\prod}}E_{i\,'}^2+\overset{N}{\underset{i_1<\,i_2}{\sum}}V_{i_1}V_{i_2}\overset{N}{\underset{i\,'\neq i_1,\,i_2}{\prod}}E_{i\,'}^2 +...+V_1V_2 \cdot \cdot \cdot V_N}{R} \nonumber \\
&=&
\frac{1}{R} \sum_{i \in {\cal N}}  V_i \prod_{ j \in {\cal N} \setminus \{ i \} }  E_{j}^2
+ \sum_{k=2}^{N} \left[ \frac{1}{R} \sum_{ {\cal C} \in {{\cal N}\choose{k}} }
 \prod_{i \in {\cal C} } V_i \prod_{j \in {\cal N} \setminus {\cal C} } E_j^2 \right], \nonumber
\end{eqnarray}
and
\begin{eqnarray}
\label{var2good}
Var(\widehat{I}_{M})
%&=&\frac{\overset{N}{\underset{i=1}{\sum}} V_i\overset{N}{\underset{i\,'\neq i} {\prod}}E_{i\,'}^2}{R}+\frac{\overset{N}{\underset{i_1<\,i_2}{\sum}}V_{i_1}V_{i_2}\overset{N}
%{\underset{i\,'\neq i_1,\,i_2}{\prod}}E_{i\,'}^2}{R^2} +...+\frac{V_1V_2 \cdot \cdot \cdot V_N}{R^N}.\nonumber \\
 &=&
\frac{1}{R} \sum_{ i \in {\cal N} }  V_i \prod_{j \in {\cal N} \setminus \{ i \}  }^N E_{j}^2
+  \sum_{k=2}^{N} \left[ \frac{1}{R^k}  \sum_{ {\cal C} \in {{\cal N}\choose{k}} }
\prod_{i \in {\cal C} } V_i \prod_{j \in {\cal N} \setminus {\cal C} } E_j^2 \right], \nonumber
\end{eqnarray}

\end{lemma}

In each case, the associated MCE  equals the square root of the corresponding variance
in Lemma \ref{prop:EstVars}. The variances (and therefore the MCEs) are asymptotically equivalent, since both converge
to zero with rate of order $\cal{O}$$(R^{-1})$. However, with the exception of the first term in
$Var(\widehat{I}_{M})$, the rest of the components in the summation converge faster to zero with rates $\cal{O}$$(R^{-k})$
for any $k \ge 2$. Hence, in finite settings the joint estimator will
always have larger error. The factors that influence the magnitude of this difference are discussed in the
next section.

\subsection{\textbf{Determinants of Monte Carlo error difference}}
\label{subsec:factorsMCEs}

In this section, we study the difference in the errors associated with
the joint and marginal estimators.
We illustrate how it depends on the dimensionality of the problem at hand ($N$),
the variation of the variables involved and the simulated
sample's size ($R$).

To begin with, if both estimators  $\widehat{I}_{J}$ and
$\widehat{I}_{M}$ are applied with the same finite $R$, then according
 to Lemma  \ref{prop:EstVars},  the difference in their variances is given by
\begin{eqnarray}\label{difvar}
Var(\widehat{I}_{J})-Var(\widehat{I}_{M})
%\hspace{-0.7em} &=&\hspace{-0.7em}
%\frac{\overset{N}{\underset{i_1<\,i_2}{\sum}}V_{i_1}V_{i_2}\overset{N}{\underset{i\,'\neq i_1,\,i_2}{\prod}}E_{i\,'}^2}{R}\left(1-\frac{1}{R}\right)
%+...+\frac{V_1V_2 \cdot \cdot \cdot V_N}{R}\left(1-\frac{1}{R^{N-1}}\right). \\
&=& \frac{1}{R}
 \sum_{k=2}^{N} \left[ \left( 1 - \frac{1}{R^{k-1}} \right)  \sum_{ {\cal C} \in {{\cal N}\choose{k}} }
\prod_{i \in {\cal C} } V_i \prod_{j \in {\cal N} \setminus {\cal C} } E_j^2 \right], \nonumber
\end{eqnarray}
\noi As the number of the variables increases, more positive terms are added to (\ref{difvar}) and
this explains the indirect effect of the dimensionality. The effect of the moments $E_i$ and
$V_i,~i=1\ldots N$, can be expressed in terms of the corresponding coefficients
of variation (CV$^2_i$), according to the following lemma.

\begin{lemma}
 \label{lemma:varMJwithCV}
Without loss of generality, let $\{ Y_i, i \in {\cal N}_0 \}$ be the sub-set of $\{Y_1, Y_2, \dots, Y_N \}$ random variable
with zero expectations.
The  variances
of the joint (\ref{JointEst}) and marginal (\ref{MargEst}) estimators are given by:

\begin{equation}\label{GivarsJ}
Var(\widehat{I}_{J})=   \frac{1}{R} \times  \prod _ {i \in {\cal N}_0} V_i~
                         \times \prod _ {i \in \overline{{\cal N}}_0} E^2_i
                         \times \left (\prod _ {i \in \overline{{\cal N}}_0} (CV^2_i+1) - I( {\cal N}_0 = \emptyset ) \right )
\nonumber
\end{equation}
and
\begin{equation}\label{GivarsM}
Var(\widehat{I}_{M})=   \frac{1}{R^{_{N_0}}}
												\times  \prod _ {i \in {\cal N}_0} V_i~
                        \times  \prod _ {i \in \overline{\cal N}_0} E^2_i
                        \times  \left (  \prod _ {i \in \overline{\cal N}_0} \left( \frac{CV^2_i}{R}+1 \right) -I( {\cal N}_0 = \emptyset ) \right )
\nonumber
\end{equation}
\noindent
where ${\cal N}_0 \subseteq {\cal N}=\{0,1,...,N\}$,
$\overline{\cal N}_0 = {\cal N} \setminus {\cal N}_0$ is the index of variables $Y_i$ with non-zero expectations,
$\prod \limits _{i \in \emptyset} Q_i = 1 $ for any $Q_i$ %(the product of the elements of an empty set is defined to be equal to one),
and  $I({\cal N}_0 = \emptyset)$ is equal to one if $E_i \neq 0$ for all $i \in {\cal N}$ %all variables have non-zero expectations
and zero otherwise.
\end{lemma}

\noi$\triangleright$
The proof of Lemma \ref{lemma:varMJwithCV} is given at the Appendix. \hfill $\Box$

\vspace{1em}
\noi Based on Lemma \ref{lemma:varMJwithCV}, the difference in the variances of the estimators
becomes larger as the variability of the $Y_i$s increases. The maximum difference occurs when
all variables involved have zero means, in which case $Var(\widehat{I}_{J})=   R^N \,Var(\widehat{I}_{M}) $.
On the contrary, when all means are non zero, the difference mainly depends on the coefficients of variation.
Based on Lemma \ref{lemma:varMJwithCV}, we may also consider the case where the two estimators have the
same variance, that is $Var(\widehat{I}_{J})=Var(\widehat{I}_{M})$,
which can be achieved under different number of replications, R$_J$ and R$_M$.
The number of replications that the joint estimator requires, in order
to archive the same error with the marginal estimator, is defined at the following corollary.

\begin{corollary}
\label{col_sameR}
The joint (\ref{JointEst}) and marginal (\ref{MargEst})   estimators achieve the same accuracy   when
    \begin{eqnarray}\label{analogy}
                    R_J &=&  R_M^{_{N_0}} \times \omega(N,N_0, {\cal CV}) \nonumber
                      \end{eqnarray}
with
$$
\omega(N,N_0,{\cal CV}) =
 \begin{cases}
 R_M^{N-N_0} & \mbox{~if~} {\cal N}_0 = {\cal N} \\
 \frac{\prod \limits_{i=1}^N (CV_i^2+1)-1}{\prod \limits_{i=1}^N (CV_i^2/R_M+1)-1} & \mbox{~if~} {\cal N}_0 = \emptyset \\
 \prod \limits_{i \in \overline{\cal N}_0} \frac{CV_i^2+1}{CV_i^2/R_M+1} & otherwise \\
 \end{cases}
$$

\noi where $N_0 = | {\cal N}_0 |$ denotes the number of the zero mean variables, ${\cal CV} = \{ CV_i: i \in \overline{\cal N}_0\}$ and
$R_J$, $R_M$ are the number of iterations for the joint and marginal estimators, respectively.
\end{corollary}

\noi  Corollary \ref{col_sameR} states that the joint estimator
achieves the same MCE when its number of iterations $R_J$ is equal to the
number of iterations of the marginal estimator $R_M$ raised to the number of variables with zero expectations
and multiplied by a factor $\omega(N,N_0,{\cal CV}) > 1$ for $R_M>1$.
Hence, in order to achieve the same precision for the two estimations,
the joint estimator will always require more iterations $R_J$ than the marginal one $R_M$.
The multiplicative factor $\omega$ heavily depends on the number of variable with zero expectations and
on the variability of the $Y_i$s (through CVs) for the non-zero variables.
In the special case where all expectations $E_i$ are  zero, the required number of iterations is R$_J$=R$_M^N$.
Lemma \ref{lemma:varMJwithCV} and  Corollary \ref{col_sameR} indicate that the error of the joint
estimator may not be always manageable. That is, if the number of variables is large or if their variability is high,
then the joint estimator requires simulated samples that can be unreasonably large.

For illustration purposes, we implement a toy example of $N$ independent and identically
 distributed (i.i.d) Beta random variables $Y_i \sim Beta(\lambda_{1},\lambda_{2})$ ($i=1,...,N$).
The mean of their product is given by:
\begin{equation}
\label{truebetameans}
E \left(\prod_{i=1}^N  Y_i \right)
 =  \left(\frac{\lambda_{1 }}{\lambda_{1 }+\lambda_{2 }}\right)^N~.\nonumber
\end{equation}

\begin{figure}[hp]
  \begin{center}\vspace{1cm}
   {\subfigure[N=10]{\label{fig:BETA_a}
                                 \psfrag{a}[c][c][0.7]{\sf Estimation of the mean (log scale)}
                                 \includegraphics[width=0.32\textwidth ]{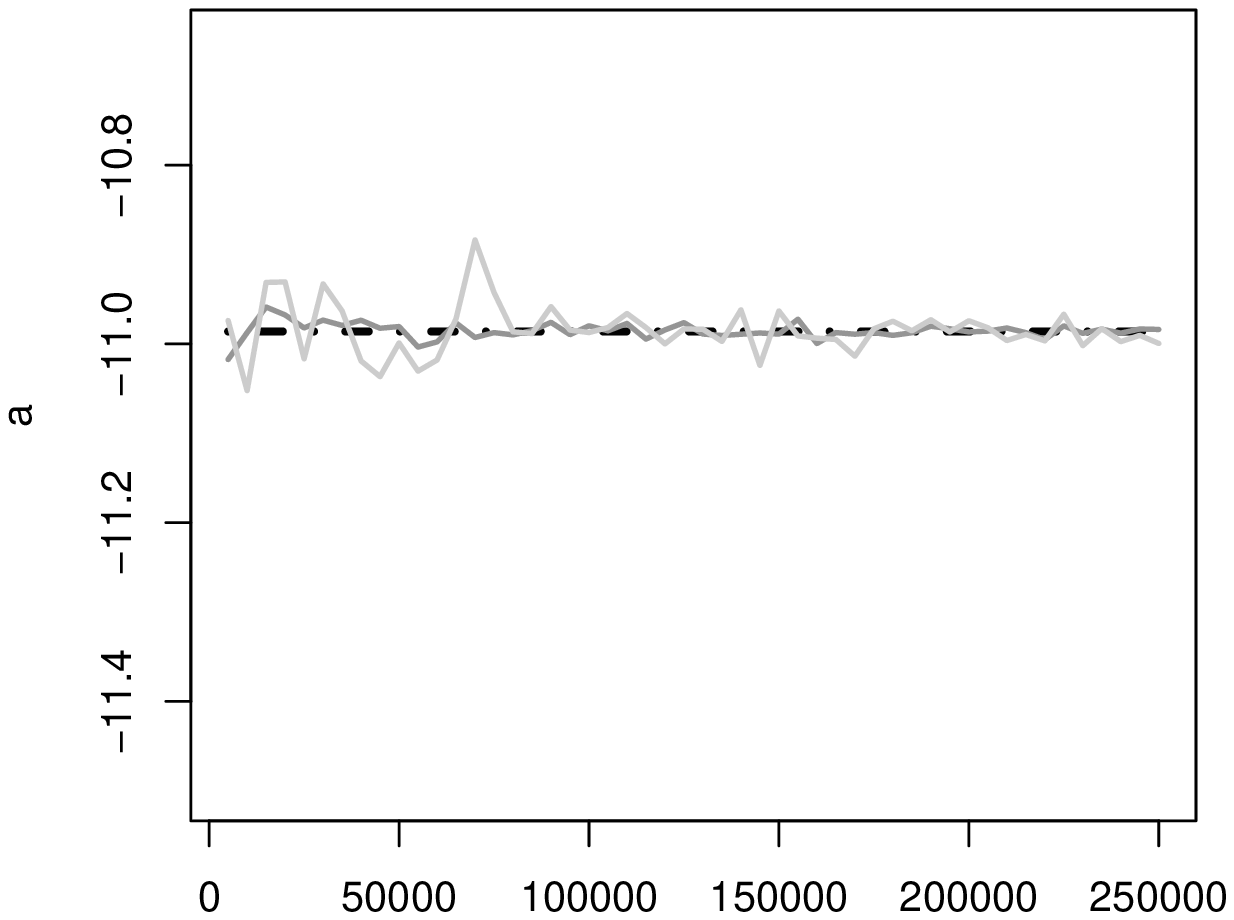}}}
   {\subfigure[N=50]{\label{fig:BETA_b}
                               \psfrag{a}[c][c][0.9]{\sf Product of N Beta(1,2) i.i.d random variables (CV=0.7)}
                                \psfrag{b}[c][c][0.8]{\sf Simulated sample size R}
                                \includegraphics[width=0.32\textwidth ]{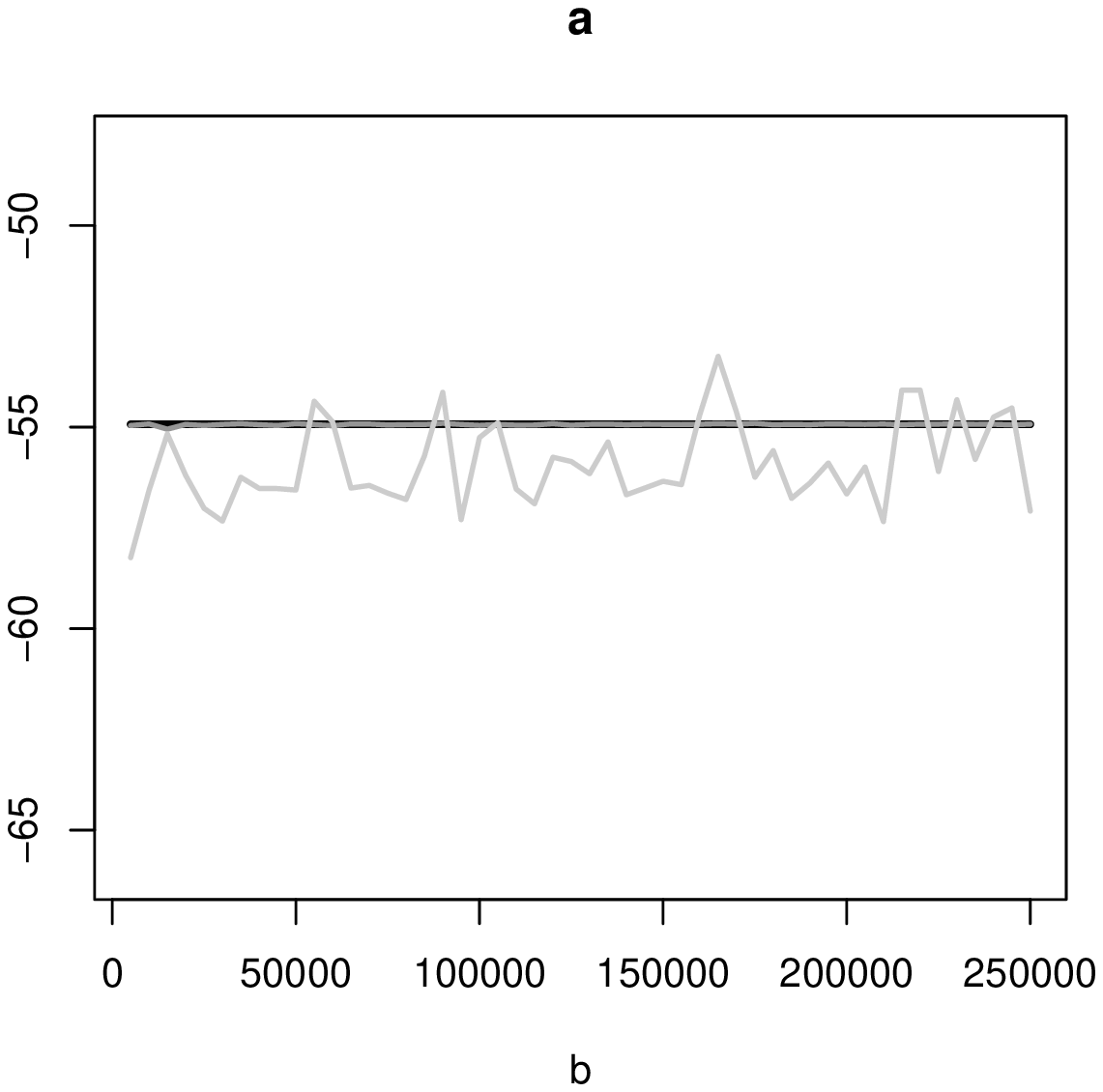}}}
  {\subfigure[N=150]{\label{fig:BETA_c}
                                 \includegraphics[width=0.32\textwidth ]{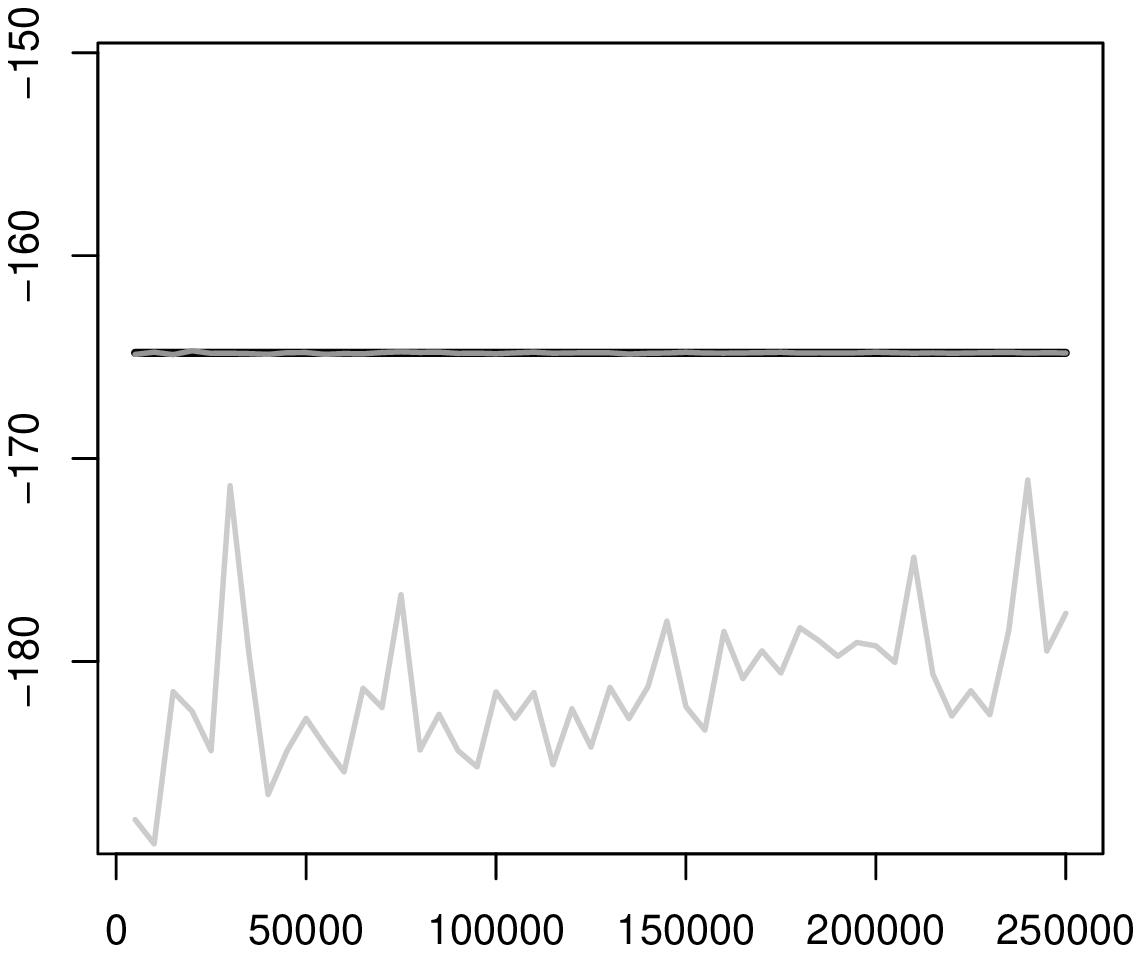}}}

{\subfigure[N=10 ]{\label{fig:BETA_d}
                                 \psfrag{a}[c][c][0.7]{\sf Estimation of the mean (log scale)}
                                 \includegraphics[width=0.32\textwidth ]{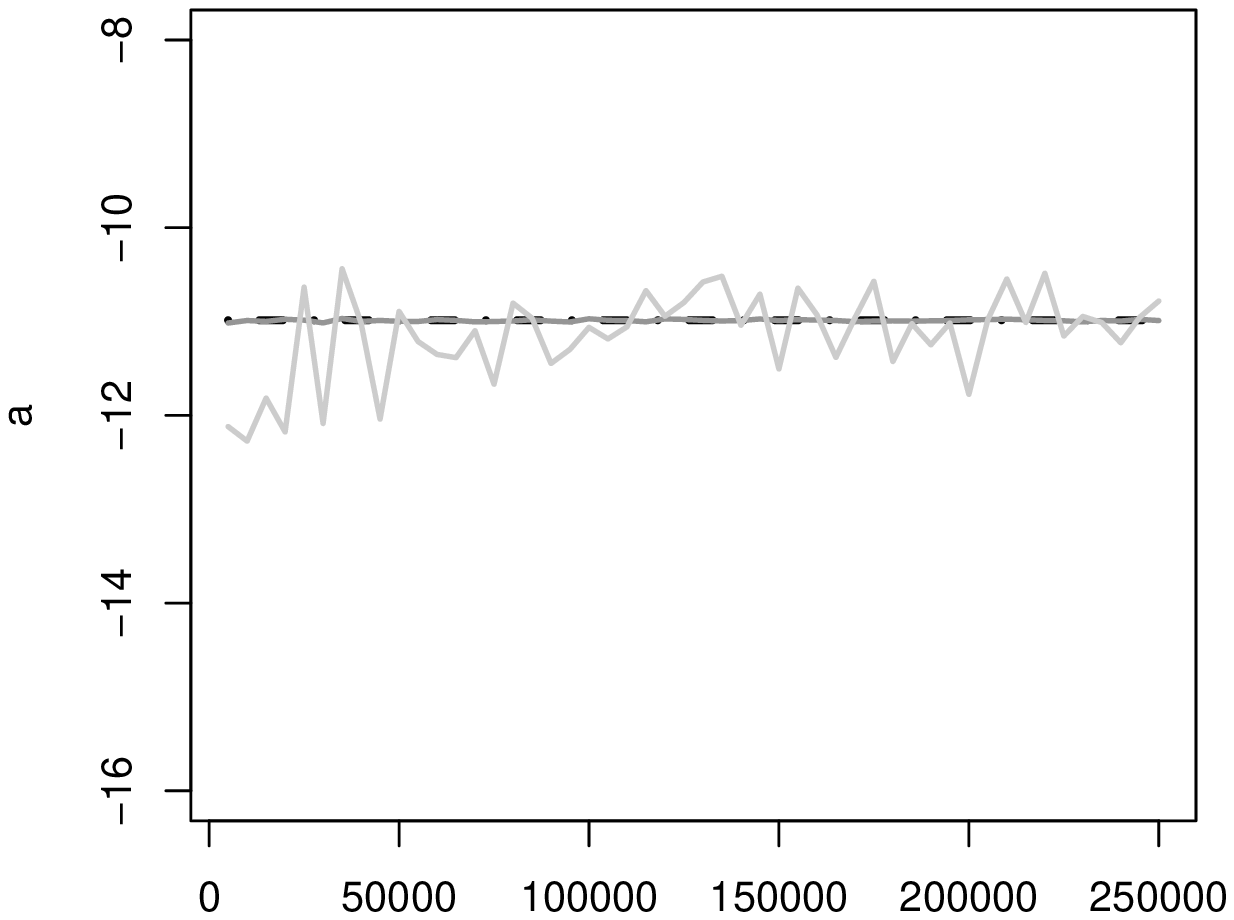}}}
{\subfigure[N=50 ]{\label{fig:BETA_e}
                               \psfrag{a}[c][c][0.9]{\sf Product of N Beta(0.1,0.2) i.i.d random variables (CV=1.24)}
                                \psfrag{b}[c][c][0.8]{\sf Simulated sample size R}
                                \includegraphics[width=0.32\textwidth ]{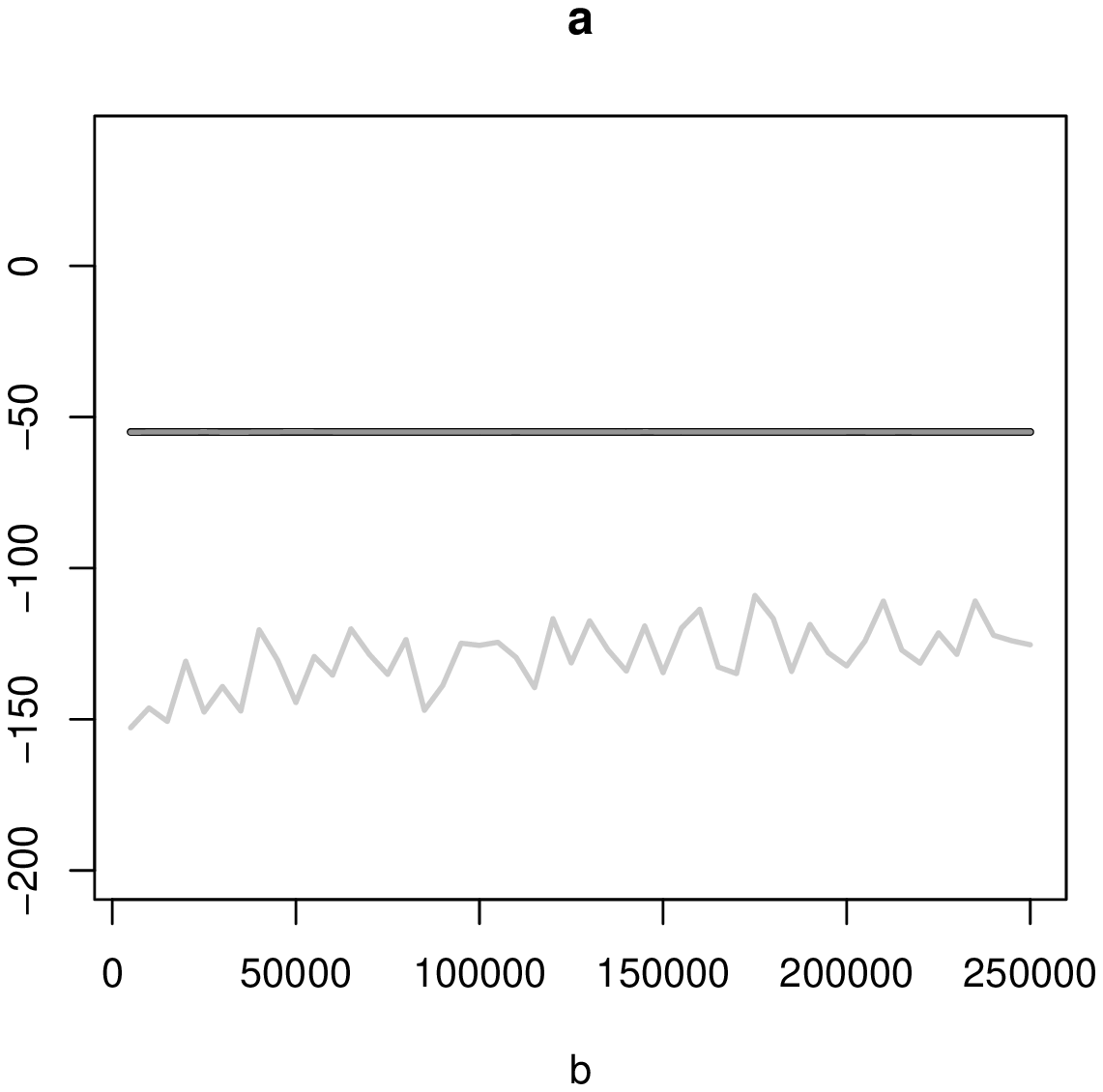}}}
{\subfigure[N=150 ]{\label{fig:BETA_f}
                                 \includegraphics[width=0.32\textwidth ]{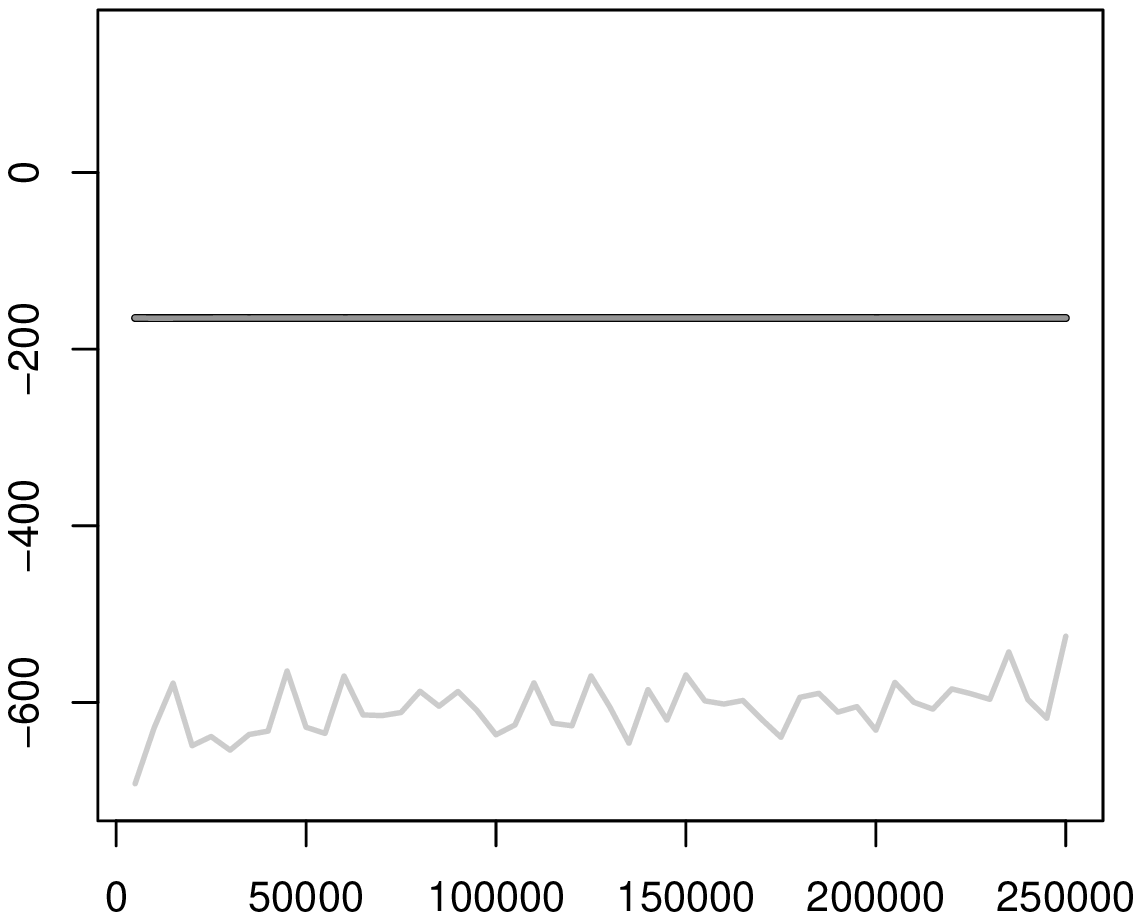}}}

    \end{center}
\caption{\footnotesize{The joint estimator $\widehat{I}_{J}$ (light grey solid line) and the marginal estimator $\widehat{I}_{M}$ (dark grey solid line) compared with the true mean (dashed black line)
of the product of $N$ i.i.d $Beta (\lambda_{1},\lambda_{2})$ variables, as the size of simulated the samples increases from 5000 to 250000 and for
$N=20, 50,$ and $150$.}}
\label{fig:batch}
\end{figure}

\noi Fifty samples with size ranging from 5 to 250 thousands simulated points,
were generated from $N=10$  $Beta(1,2)$ distributions.
The two estimators were computed and  depicted in Figure \ref{fig:BETA_a}.
The same procedure was repeated for $N=50$ and $N=150$ and
is graphically represented in Figures \ref{fig:BETA_b} and \ref{fig:BETA_c}.

In the low dimensional case ($N=10$), the error of the joint estimator ($\widehat{I}_{J}$: light grey line)
is rather comparable with the error of marginal ($\widehat{I}_{M}$: dark grey line).
When $R$ reaches 250 thousands, both estimators reach the true mean ($I_{T}$: dashed line).
However, if the number of variables is increased to $N=50$ and $N=150$,
the variability differences between the two approaches remain large even for $R=250,000$;
see Table \ref{table:beta_margjoint}.

The exercise was also replicated for $N=10, 50$ and $150$ i.i.d. $Beta(0.1,0.2)$ variables.
The true mean is the same with the previous setting (equal to $1/3$), but the coefficient of variation (CV) is
now approximately $77\%$ higher. For the same $R$ and $N$, the difference in the errors of
the two estimators is even larger (Figures \ref{fig:BETA_d} to \ref{fig:BETA_e}), indicating
the role of the variability of the variables involved. The estimated values and the
corresponding errors are summarized in Table \ref{table:beta_margjoint}.
Although, this example is simple assuming i.i.d random variables,
the same picture can be reproduced for non identically distributed random variables.

\begin{table}[h!]
\caption{\small \textbf{Estimated mean of the product of i.i.d Beta variables (log scale)}}
\label{table:beta_margjoint}
\begin{center}
\begin{tabular} {llrrrrr}
       \\
Distribution                         & N  & $I_T$ & $\widehat{I}_{M}$  & $M\widehat{C}E_{M}$ & $\widehat{I}_{J}$   &  $M\widehat{C}E_{J}$  \\
\hline
\hline
\multirow{3}{*}{$Beta(1,2)$}        & 10   &  -10.99 &  -10.98  & 0.02 &  -10.97 &  0.07 \\
                                    & 50   &  -54.93 &  -54.93  & 0.06 &  -52.01 &  2.03 \\
                                    & 150  & -164.79 & -164.79  & 0.09 & -176.94 &  3.37 \\
\\
\multirow{3}{*}{$Beta(0.1,0.2)$}    & 10   &  -10.99 &  -10.98 & 0.04 &  -11.05 &   1.07  \\
                                    & 50   &  -54.93 &  -54.90 & 0.10 & -113.81 &  13.77 \\
                                    & 150  & -164.79 & -164.80 & 0.17 & -595.13 &  28.50 \\

 \hline
\multicolumn{7}{p{12.4cm}}{\scriptsize
       $N$: Number of i.i.d variables;
       $I_T$: true mean;
       $\widehat{I}_{(J\,or\,M)}$: the estimated value via the joint or the marginal approach respectively, over $R=$250,000 iterations;
       $\widehat{I}_{(M\,or\,J)}$ and $M\widehat{C}E_{(M\,or\,J)}$: batch mean error over 25 batches of 10,000 points each
       (obtained as the standard deviation of the log estimates).
        }

\end{tabular}
\end{center}
\end{table}

\newpage
\subsection{\textbf{Variance reduction under conditional independence}} \label{subsec:Cond_MCEs}
\label{subsub_MCEfactors}

In this section, we demonstrate how we can extend the previous results in the case of conditional independence
which is more realistic in practice and it frequently met in hierarchical models with latent variables.

Specifically, let us substitute $\dn{Y}$ by $(\textbf{\emph{U}},\textbf{\emph{V}}$).
In analogy with the previous setting, let $\dn{U}_i$ (with $i=1,2,\dots,N$)
be conditionally independent random variables when $\dn{V}$ are given
with densities denoted by $h( \dn{u}_i | \dn{v} )$.
We are interested in estimating the integral
\begin{equation}
\label{cond1}
\mathcal{I}=\int \Big[ \prod_{i=1}^{N}  \varphi_i(\textbf{\emph{u}}_i,\textbf{\emph{v}}) \Big]
                                              h(\textbf{\emph{u}},\textbf{\emph{v}})\,d(\textbf{\emph{u}},\textbf{\emph{v}}),
\end{equation}
\noi that now corresponds to the expected value of
$\varphi(\textbf{\emph{u}},\textbf{\emph{v}}) = \prod_{i=1}^{N}  \varphi_i(\textbf{\emph{u}}_i,\textbf{\emph{v}})$ over $h(\textbf{\emph{u}},\textbf{\emph{v}})$.
This can be directly estimated by the joint estimator
\begin{equation}
\widehat{\mathcal{I}}_J = \frac{1}{R} \sum_{r=1}^{R} \left[
													\prod_{i=1}^{N}  \varphi_i \big(\textbf{\emph{u}}_i^{(r)},\textbf{\emph{v}}^{(r)}\big) \right]
\label{joint_cond_ind}
\end{equation}
assuming that we can generate a random sample $\big\{ \dn{u}^{(r)}, \dn{v}^{(r)}\big\}_{r=1}^R$ from $h(\dn{u},\dn{v})$.

If we use the conditional independence assumption, (\ref{cond1}) can be  written as
\begin{equation}
\label{cond3}
\mathcal{I} = \int \left\{  \prod_{i=1}^N \Big[
               \int \varphi_i(\textbf{\emph{u}}_i,\textbf{\emph{v}})  h(\textbf{\emph{u}}_i|\textbf{\emph{v}})\,d \emph{\textbf{u}}_j \Big]  \right\}                        h(\textbf{\emph{v}})\,d \emph{\textbf{v}}
              = \int \prod_{i=1}^N E\big( \varphi_i \big|\textbf{\emph{v}} \big) h(\textbf{\emph{v}})\,d \emph{\textbf{v}},
\end{equation}
where $ E\big( \varphi_i \big|\textbf{\emph{v}} \big)$ is the conditional expectation of $\varphi_i( \dn{u}_i, \dn{v} )$
with respect to $h( \dn{u}_i | \dn{v} )$.
From (\ref{cond3}) we can directly obtain the corresponding marginal estimator by
\begin{equation}
\widehat{\mathcal{I}}_M = \frac{1}{R_1} \sum_{r_1=1}^{R_1} \Big[ \prod_{i=1}^N \overline{\varphi}_i^{(r_1)} \Big]
\mbox{~~with~~}
\overline{\varphi}_i^{(r_1)} = \frac{1}{R_2} \sum_{r_2=1}^{R_2} \varphi_i \big( u_i^{(r_2)}, \dn{v}^{(r_1)} \big),
\label{marg_cond_ind}
\end{equation}
calculated by a nested Monte Carlo experiment;
where $\big\{ \dn{v}^{(r_1)} \big\}_{r_1=1}^{R_1}$ is a sample from $h(\dn{v})$ and
$\big\{  u_i^{(r_2)} \big\}_{r_2=1}^{R_2}$ is a sample obtained by the conditional distribution
$h\big( u_i | \dn{v}=\dn{v}^{(r_1)} \big)$.

% ---
\begin{lemma}
\label{var_cond_ind}
The  variances of the joint (\ref{joint_cond_ind}) and  marginal estimators (\ref{marg_cond_ind})
under the assumption of conditional independence are given by
\begin{eqnarray*}
\label{var1good_cond}
Var(\widehat{I}_{J})
&=&
\frac{1}{R} Var_{\dn{v}} \Big[ \prod_{i=1}^N E\big( \varphi_i \big| \dn{v} \big)  \Big]
+\frac{1}{R}
\sum_{k=1}^{N} \sum_{ {\cal C} \in {{\cal N}\choose{k}} }
 E_{\dn{v}}\Big[ \prod_{i \in {\cal C} } V\big( \varphi_i \big| \dn{v} \big) \prod_{j \in {\cal N} \setminus {\cal C} } E\big( \varphi_j \big| \dn{v} \big)^2 \Big]
\end{eqnarray*}
and
\begin{eqnarray*}
\label{var2good_cond}
Var(\widehat{I}_{M})
&=&
\frac{1}{R_1} Var_{\dn{v}} \Big[ \prod_{i=1}^N E\big( \varphi_i \big| \dn{v} \big)  \Big]
+\frac{1}{R_1}
\sum_{k=1}^{N} \frac{1}{R_2^k}
\sum_{ {\cal C} \in {{\cal N}\choose{k}} }
 E_{\dn{v}}\Big[ \prod_{i \in {\cal C} } V\big( \varphi_i \big| \dn{v} \big) \prod_{j \in {\cal N} \setminus {\cal C} } E\big( \varphi_j \big| \dn{v} \big)^2 \Big]
\end{eqnarray*}
where $E_{\dn{v}} \big[ g(\dn{v}) \big]$ and $Var_{\dn{v}} \big[g(\dn{v}) \big]$ denote the expectation and the variance of $g(\dn{v})$
with respect to $h(\dn{v})$  and $V\big( \varphi_i \big| \dn{v} \big)$ is, in analogy to  $E\big( \varphi_i \big| \dn{v} \big)$, the conditional variance of $\varphi_i( \dn{u}_i, \dn{v} )$ with respect to $h( \dn{u}_i | \dn{v} )$.

\end{lemma}

\noi$\triangleright$
The proof of Lemma \ref{var_cond_ind} is given at the Appendix. \hfill $\Box$

\vspace{1em}

Lemma \ref{var_cond_ind} is an extension of Lemma \ref{prop:EstVars} for the case of conditional independence.
For this reason, similar statements
about the behaviour and the error of the joint and the marginal estimators
also hold for the case of conditional independence.
The main difference is the first term of variances of the estimators which is common and it is due
to the additional variability of $\dn{v}$ which is of order ${\cal O}(R^{-1})$.
Moreover, for $R_1=R$ and any $R_2>1$ the marginal estimator is better since $Var(\widehat{I}_{M})<Var(\widehat{I}_{J})$.
It would be interesting to examine the case
of using the exactly the same computation effort in terms of Monte Carlo iterations.
Nevertheless, setting $R=R_1R_2$, then no clear conclusion can be drawn since the
first common term will be of different order.
For example, if we consider $R_1=R_2=r$ and $R=r^2$ then the two variances are given by
\begin{eqnarray*}
\label{var1good2}
Var(\widehat{I}_{J})
&=&
\frac{1}{r^2} Var_{\dn{v}} \Big[ \prod_{i=1}^N E\big( \varphi_i \big| \dn{v} \big)  \Big]
+\frac{1}{r^2}
 \sum _{i =1 }^N E_{\dn{v}}\Big[  V\big( \varphi_i \big| \dn{v} \big)
 \prod_{j \in {\cal N} \setminus \{ i \} } \hspace{-0.7em} E\big( \varphi_j \big| \dn{v} \big)^2 \Big] + {\cal O}(r^{-2})
\end{eqnarray*}
and
\begin{eqnarray*}
\label{var2good2}
Var(\widehat{I}_{M})
&=&
\frac{1}{r} Var_{\dn{v}} \Big[ \prod_{i=1}^N E\big( \varphi_i \big| \dn{v} \big)  \Big]
+\frac{1}{r^2 }
 \sum _{i =1 }^N E_{\dn{v}}\Big[  V\big( \varphi_i \big| \dn{v} \big)
 \prod_{j \in {\cal N} \setminus \{ i \} } \hspace{-0.7em} E\big( \varphi_j \big| \dn{v} \big)^2 \Big] + {\cal O}(r^{-3})
\end{eqnarray*}
\noi
Finally, in the case that instead of nested Monte Carlo, we use a numerical method
which approximates very well the expectations $E( \varphi_i | \dn{v} )$
then  the second term of the the variance of the corresponding marginal estimator will be zero making
the method considerably more accurate and faster to converge than the joint estimator.

Due to the fact that Lemma \ref{var_cond_ind} also incorporates similar expressions as in Lemma \ref{lemma:varMJwithCV},
the remarks made on the error differences with regard to the sample size, the number of variables and their
variability apply also in the case of  conditional independence assumption. We may now explain the different
behaviour of the three BML estimators at the GLLVM
example (Section \ref{sec_gllvm}),
where $\dn{u}_i = \dn{Z}_i$ are the latent variables and $\dn{v}=(\dn{\alpha}, \dn{\beta})$ are the model parameters.
The error differences observed in Figure \ref{fig:bridge_a}
between the $BH_J$ and $BG_J$  estimators (for the same $N$ and $R$)
can be now attributed to the
different coefficients of variation of the averaged quantities involved. For both estimators, the expectation in the nominator
is taken over $g(\dn{\alpha}, \dn{\beta},\dn{Z}) = g(\dn{\alpha})g(\dn{\beta}) \prod_{i=1}^N (\dn{Z}_i)$. However, the $N$ averaged variables differ according to
(\ref{harm}) and (\ref{geom}).
Specifically for $i=1,\dots,N$ the averaged variables were:
\begin{itemize}
    \item[] (a) $\varphi_i(\cdot)=\Big[ g(\dn{\alpha})^{1/N} g(\dn{\beta})^{1/N} g(\textbf{Z}_i) \Big]^{-1}$, in the case of  BH$_J$ and
    \item[] (b) $\varphi'_i(\cdot)=\
    \left\{        \tfrac{ f(Y_i|\dn{\alpha}, \dn{\beta},\dn{Z}_i) \pi(\dn{Z}_i)}{ g(\dn{Z}_i) }
            \left[ \tfrac{ \pi(\dn{\alpha})  \pi(\dn{\beta}) }{ g(\dn{\alpha}) g(\dn{\beta})}\right]^{1/N}
            \right\}^{1/2}$,
    in the case of BG$_J$.
\end{itemize}
Moreover, none of the conditional expectations will be equal to zero since $\phi_i$ and $\phi_i'$ are both positive.
Therefore, following Lemma \ref{lemma:varMJwithCV} we may rewrite the variances of the estimators as functions
of the corresponding coefficients of variation
\begin{eqnarray*}
Var(\widehat{I}_{J})
\hspace{-0.7em} &=& \hspace{-0.7em}
\frac{1}{R} Var_{\dn{v}} \Big[ \prod_{i=1}^N E\big( \varphi_i \big| \dn{v} \big)  \Big]
+\frac{1}{R}    E_{\dn{v}}\left[ \prod _ {i =1}^N E\big( \varphi_i \big| \dn{v} \big)^2
                \Bigg\{ \prod _ {i =1}^N \Big[ CV\!\left( \varphi_i | \dn{v} \right)^2 + 1 \Big] - 1  \Bigg\} \right]
\end{eqnarray*}
and
\begin{eqnarray*}
Var(\widehat{I}_{M})
\hspace{-0.7em} &=& \hspace{-0.7em}
\frac{1}{R_1} Var_{\dn{v}} \Big[ \prod_{i=1}^N E\big( \varphi_i \big| \dn{v} \big)  \Big]
+\frac{1}{R_1}
   E_{\dn{v}}\left[         \prod _ {i =1}^N E\big( \varphi_i \big| \dn{v} \big)^2
                    \Bigg\{ \prod _ {i =1}^N \Big[ \tfrac{CV\left( \varphi_i | \dn{v} \right)^2 }{R_2}  +1 \Big]
                    -1  \Bigg\} \right]
\end{eqnarray*}
From the above equations, it is obvious that the variances of the estimators will explode for large $N$
in the (a) case since we expect values of $\varphi_i > 1$ demanding a large number of iterations
to reach a required precision level. The effect will be more evident in the joint estimator,
since the marginal estimator some of these effects will be eliminated for large $R_2$
(or using well behaved numerical methods).
For case (b), the situation seems much better, since (assuming that $g$ is a good proxy for the posterior)
the expectation in the first term (which is common in both approaches) will estimate the normalizing constant
of $f(\dn{\alpha}, \dn{\beta} | \dn{y})$ for given values of $\dn{\alpha}$ and $\dn{\beta}$.
These values are usually small and therefore will not to greatly influenced by $N$.
Therefore this term will be eliminated for reasonably small $R$ and $R_1$.
If this is the case, the second term will behave as in described in previous sections and therefore
any action of marginalizing will greatly improve the Monte Carlo errors.

To verify this, we used the last 5000 iterations to calculate the corresponding $CV$s. For the bridge harmonic estimator,
the $CV$s of the $N$ quantities in (a) varied in log scale from 0.20 to 0.52 (median $CV$=0.27).
In  the case of the bridge geometric estimator, the $CV$s of the corresponding variables in (b) were substantially lower,
varying from 0.01 to 0.10 (median $CV$=0.02). Similar results occurred for the denominators of the two bridge sampling estimators
(harmonic: $CV$ from 0.2 to 0.9 /geometric: $CV$ less than 0.006).

The conditional independence setting considered here, applies to a plethora of high dimensional models involving latent vectors
and it provides formally the rational behind choosing to marginalize out the latent variables.
In such settings, the rate of convergence is extremely slow and millions of iterations may be
required to achieve a desirable level of precision for the joint estimator.
However,  convergence is not only a matter of the associated MCE, as will be explained in the next section.

\bigskip
\subsection{\textbf{The role of the sample covariation}} \label{sec:OnCov}
Up to this point, we have studied the variability differences between the two approaches under consideration.
In this section, we focus on the estimators themselves and how they are influenced by
sample covariation which are expected to be close (but not exactly) equal to zero.
These differences appear in the simulated example of Section \ref{subsec:simGLLVM}
(see Tables \ref{table:GLLVM_margjoint} and \ref{table:beta_margjoint}
and cannot be attributed to the associated Monte Carlo errors of the two estimators.
In the bivariate case, the difference between the mean of the product of two variables and the product of their means
is by definition their covariance. Let us refer to a multivariate analogue of covariance with the general
term \emph{total covariation} defined as:

\begin{equation}
\label{covariation}
TCI( \dn{Y} )=E\Big(\prod_{i=1}^N Y_i\Big)-\prod_{i=1}^NE(Y_i),
\end{equation}

%E\left[\prod_{i=1}^N\phi_i(Y_i)\right]

\noi which is actually the difference between the expectations under the joint and marginal approaches in their simplest forms.
For instance, it coincides with the difference between the expressions  in (\ref{basicexpJoint})
and (\ref {basicexpMarg}) if in (\ref{covariation}) we use  the random variables $\phi_i(Y_i),~i=1,...,N$ (for simplicity
in the notation hereafter we proceed with the original variables without loss of generality).
The identity (\ref{covariation}) is not useful into gaining insight on the factors that affect
that difference. Here, we provide an alternative expression which assesses the total covariation among $N$ random variables,
in terms of their expected means $E(Y_i),\, i=1,...,N$ and covariances of the form:
\begin{equation}
\label{covN}
Cov_{(k)}(\dn{Y})=Cov\Big(\overset{k-1}{\underset{i=1}{\prod}}Y_i,Y_k\Big).
\end{equation}

\begin{lemma}
\label{lemma:ISDI}
The total covariation among N variables, is given by:
\begin{equation}\label{isdi1a}
\mbox{TCI}(\dn{Y})
=Cov_{(N)}(\dn{Y}) + \sum_{k=1}^{N-2} \left[ \left( \prod^{N }_{i=N-k+1} \!\!\!\! E (Y_i) \right) Cov_{(N-k)}(\dn{Y}) \right],
\end{equation}
\\where  $N\geq 3$ and $E( Y_{N+1} )=1$.
\end{lemma}

\noi$\triangleright$
The proof of Lemma \ref{lemma:ISDI} is given at the Appendix. \hfill $\Box$

The total sample covariation among the $N$ random variables is therefore assessed through a weighted sum of $N$-1 covariance terms.
The means of the variables serve as weights that adjust the contribution to the total covariation for each additional variable.
In finite settings, the difference between the estimated means provided by $\widehat{I}_{J}$ and $\widehat{I}_{M}$
reflects the total sample covariation between the $N$ variables.

When $N$ random variables are simulated independently,
even the smallest dependencies between the variables will  result in non zero total sample covariation.
That is, even though the $N$ variables were sampled independently, the covariance induced
by the simulation procedure cannot be ignored even for samples of several hundreds of
thousands points. Therefore, if the total sample covariation  is non zero, it can be considered as an
index of the sample's divergence from independence. It should be noted that zero values do not
ensure independence (that is, the reverse statement does not hold). By definition, the
total sample covariation is accountable for and completely explains the
estimation differences that were illustrated in the our examples.

Equation  (\ref{isdi1a}) implies that any divergence from the independence assumption in finite settings
is also affected by the number of variables $N$, their expectations, their covariation and the simulated sample size $R$, as already illustrated graphically
in Figures \ref{fig:BETA_a} to \ref{fig:BETA_f}. In the case of independent variables,
the sample covariation converges to zero as $R$ goes to infinity.
The Cauchy-Schwartz inequality provides an upper bound for the sample covariation, according to the following lemma.

\begin{corollary}
\label{lemma:CS_lowerbound}
An upper bound for the absolute value of $TCI(\dn{Y})$ is given by:

\begin{eqnarray}
|TCI(\dn{Y})|
%& = & \sum_{k=0}^{N-2} \left|\,  Cov\Bigg( \prod_{j=1}^{N-k-1} \!\!\! Y_i,~Y_{N-k} \Bigg)  ~~
%                                   \prod_{i=N+1-k}^{N+1} \!\!\!\!\!\! E_i ~ \right|\nonumber \\
& \leq &  \sum_{k=0}^{N-2}\left[   \Bigg( \prod_{i=N+1-k}^{N+1} \!\!\!\!\!\! \lvert\,E (Y_i) \rvert \Bigg)
 \sqrt{ Var \left(\overset{N-k-1}{\underset{j=1}{\prod}}Y_i\right) Var(Y_{N-k}) }
 ~~
 ~ \right].\nonumber
\end{eqnarray}

\end{corollary}

\noi$\triangleright$
Corollary \ref{lemma:CS_lowerbound} immediately follows from Lemma \ref{lemma:ISDI} by further implementing
the Cauchy-Schwartz inequality.
%\begin{equation}\label{c-s}
%\lvert\, Cov(Y_1,Y_2)\rvert\leqq \sqrt{Var(Y_1)\,Var(Y_2)}=sd(Y_1)sd(Y_2).\nonumber
%\end{equation}
%
\hfill $\Box$

\vspace{1em}
Corollary \ref{lemma:CS_lowerbound}  provides an upper end to the total covariation therefore we cannot infer regarding the
its magnitude as the various parameters increase. However, in a vise versa
point of view, Lemma \ref{lemma:CS_lowerbound} suggests that:

\begin{itemize}
    \item [--]  The lower the expected means of the variables (in absolute value) are, the lower the index is expected to be (due to the lower bound).
    \item [--]  The lower the variances of the variables are, the lower the index is expected to be (due to the lower bound).
    \item [--] Less variables (smaller $N$) correspond to lower number of positive terms added to the right part of the inequality
    and therefore to lower total covariation.
\end{itemize}

The total sample covariation affects also the estimated variance of the joint estimator.
Let us denote with $R_0$, the number of iterations required to overcome the sample covariation effect. For
simulated samples less that $R_0$, the variance of the joint estimator is underestimated by a
factor of $TCI(\dn{Y})^2$, according to the following lemma.

 \begin{lemma}
 \label{lemma:underestimaton}
The variance of the product of $N$ variables, equals  their
variance under assumed independence  minus the square of their total covariation,
\begin{eqnarray}
  Var \left(  \prod_{i=1}^N Y_i \right) = Var \left(  \prod_{i=1}^N Y_i \Big| \mbox{Independence} \right) - TCI(\dn{Y})^2~,
\end{eqnarray}
where $Var \left(  \prod_{i=1}^N Y_i \Big| \mbox{Independence} \right) $ is the variance of the product under
the assumption of independence.
\end{lemma}

\noi$\triangleright$
The proof of Lemma \ref{lemma:underestimaton} is given at the Appendix. \hfill $\Box$

\vspace{1em}
\noi According to Lemma \ref{lemma:underestimaton}, in the presence of sample total covariation, the
joint approach leads in practice to a false sense of accuracy. Once the
simulated sample is large enough (larger than R$_0$), the covariation effect vanishes ($TCI(\dn{Y})^2 \simeq 0$),
 yet the variance of the joint estimator
is always larger than the one associated with the
marginal estimator, according to (\ref{difvar}).

Based on the sample total covariation of $\dn{\Phi} = \big( \phi_1(Y_1), \dots,  \phi_N(Y_N) \big)$, it is now possible to explain why
at the GLLVM example (Section \ref{sec_gllvm})  MCMC estimators associated with low MCE
lead to biased estimations and vice versa. In particular, the sample covariation
does not seem to affect the bridge harmonic (BH$_J$) estimator while it is clearly present in the case of the
reciprocal (RM$_J$) estimator (see Table \ref{table:GLLVM_margjoint}).
To explain this phenomenon, we need first to underline that the bridge harmonic estimator is a ratio.
Based on the last 5,000 draws, the sample total covariation between the averaged variables at the nominator of
BH$_J$ was -723.8 and -730.5 at the denominator. These values are substantially larger than
the sample covariation among the averaged variables in the case of the reciprocal estimator (equal to -23.0).
However, since BH$_J$ is a ratio the sample covariations estimated at the nominator and the denominator cancel out,
which is not the case for the reciprocal estimator.
Similarly, the sample covariation effect also cancels out in the case of the bridge geometric estimator.

\bigskip
\section{\textbf{Discussion}}\label{sec_disc}

In the presence of independence assumptions, the mean product of $N$ variables can be either estimated by
implementing the joint or the marginal approaches, as described in the current
work. In finite settings the difference may be considerable, making the selection of one of the approaches
crucial for the accurate estimation of specific quantities. It might seem appealing to
adopt the joint approach in order to simplify the estimator and
minimize the computational burden and the corresponding time required.
In fact, such a gain is not obtained in practice, since the joint approach
is associated with increased error and divergence from the true mean.
As discussed in Section \ref{sec:2estim} and illustrated at the examples,
the number of iterations required for the joint estimator to obtain
values close to the true mean  is considerably higher than the one required for the marginal estimator.
In complex settings, the number of iterations might be so large, that lack of convergence may remain undetected.

\newpage

\section*{APPENDIX}\label{append_A}

The identities of the MCMC estimators used in the Section \ref{subsec:model} are

\begin{itemize}

  \item \emph{Reciprocal importance (RM) sampling} estimator \citep{GelDey94}
            \begin{equation}\label{recipr}
                f(\dn{Y})= \left[ \int \frac{g(\boldsymbol{\vartheta})}{f(\textbf{Y}|\,\boldsymbol{\vartheta})\,\pi(\boldsymbol{\vartheta})} \,\pi(\boldsymbol{\vartheta}|\,\textbf{Y})\,d{\boldsymbol{\vartheta}}\right ]^{-1},
            \end{equation}

 \item \emph{Generalized harmonic bridge (BH) sampling} estimator \citep{me:wo96}
             \begin{equation}\label{harm}
                 f(\dn{Y})=\frac{  \displaystyle\int \left[\,g(\boldsymbol{\vartheta\underline{}}) \right]^{-1}g(\boldsymbol{\vartheta})\,d{\boldsymbol{\vartheta}}}
                 {\displaystyle\int\left [  f(\textbf{Y}|\,\boldsymbol{\vartheta})\pi(\boldsymbol{\vartheta})   \right ]^{- 1}\pi(\boldsymbol{\vartheta}|\,\textbf{Y})\,d{\boldsymbol{\vartheta}}}\,,
            \end{equation}

 \item  \emph{Geometric  bridge (BG) sampling} estimator \citep{me:wo96}
             \begin{equation} \label{geom}
                 f(\textbf{Y})=\frac{  \displaystyle\int \left[ \frac{ f(\textbf{Y}|\,\boldsymbol{\vartheta})\pi(\boldsymbol{\vartheta}) } {g(\boldsymbol{\vartheta}) } \right]^{1/2}g(\boldsymbol{\vartheta})\,d{\boldsymbol{\vartheta}}}
                    {\displaystyle\int \left[ \frac{ f(\textbf{Y}|\,\boldsymbol{\vartheta})\pi(\boldsymbol{\vartheta}) } {g(\boldsymbol{\vartheta}) } \right]^{-1/2}\pi(\boldsymbol{\vartheta}|\,\textbf{Y})\,d{\boldsymbol{\vartheta}}}\,\,.
             \end{equation}

\end{itemize}

\vspace{1.5cm}

\textbf{Proof of Lemma \ref{lemma:varMJwithCV} }

According to \cite{good62}, the variance of the product of N variables is given by

\begin{equation}\label{basevar}
Var\left(\prod_{i=1}^N\phi_i(Y_i)\right)
=\prod_{i=1}^N\left(V_i+E_i^2\right)-\prod_{i=1}^NE_i^2.
\end{equation}

Hence we can write
\begin{eqnarray*}
Var\left(\prod_{i=1}^N\phi_i(Y_i)\right)
&=& \prod_{i \in {\cal N}_0} \left(V_i+E_i^2\right)\prod_{i \in \overline{\cal N}_0} \left(V_i+E_i^2\right)\
     -\prod_{i \in {\cal N}_0}E_i^2 \prod_{i \in \overline{\cal N}_0}E_i^2 . \\
&=& \prod_{i \in {\cal N}_0} V_i   \prod_{i \in \overline{\cal N}_0} \Big[ E_i^2 \left(CV_i^2+1\right) \Big]
     -\prod_{i \in {\cal N}_0}E_i^2 \prod_{i \in \overline{\cal N}_0}E_i^2 . \\
&=& \prod_{i \in \overline{\cal N}_0}  E_i^2
		\times  \left[ \prod_{i \in {\cal N}_0} V_i   \prod_{i \in \overline{\cal N}_0}   \left(CV_i^2+1\right)
     -\prod_{i \in {\cal N}_0}E_i^2 \right] .
\end{eqnarray*}
Note that $\prod \limits _{i \in {\cal N}_0}E_i^2 $ will be the value of one if ${\cal N}_0 = \emptyset$ and zero otherwise.
Therefore we can write
$\prod\limits _{i \in {\cal N}_0}E_i^2 = \prod\limits _{i \in {\cal N}_0}E_i^2 \times \prod\limits _{i \in {\cal N}_0}V_i^2$
resulting in
\begin{eqnarray*}
Var\left(\prod_{i=1}^N\phi_i(Y_i)\right)
&=& \prod_{i \in {\cal N}_0} V_i \times
		\prod_{i \in \overline{\cal N}_0}  E_i^2  \times
		\left[   \prod_{i \in \overline{\cal N}_0}   \left(CV_i^2+1\right)
     -\prod_{i \in {\cal N}_0}E_i^2 \right] . \\
&=& \prod_{i \in {\cal N}_0} V_i \times
		\prod_{i \in \overline{\cal N}_0}  E_i^2  \times
		\left[   \prod_{i \in \overline{\cal N}_0}   \left(CV_i^2+1\right)
     -I( {\cal N}_0 = \emptyset ) \right], \\
\end{eqnarray*}
%where $I(A)$ is the indicator function taking the value of one if $A$ is true and zero otherwise.
which gives
$$
Var\left(\prod_{i=1}^N\phi_i(Y_i)\right) = \hspace{13cm}
$$
\begin{eqnarray*}
&=&
\left\{
\begin{array}{ll}
\prod\limits_{i=1}^N V_i  & \mbox{ if } {\cal N}_0={\cal N}  \mbox{ (all expectations are zero)}\\
% 			\multicolumn{2}{r}{ \mbox{ (all expectations are zero)} } \\[1em]
\prod\limits_{i =1}^N  E_i^2  \times   \Big[ \prod\limits_{i =1}^N   \big(CV_i^2+1 \big) -1  \Big]&
 			\mbox{ if } {\cal N}_0=\emptyset \mbox{ (all expectations are non-zero)} \\
 			%\multicolumn{2}{r}{ \mbox{ (all expectations are non-zero)} } \\[1em]
\prod\limits_{i \in {\cal N}_0} V_i \times
		\prod\limits_{i \in \overline{\cal N}_0}  E_i^2  \times
		   \prod\limits_{i \in \overline{\cal N}_0}   \left(CV_i^2+1\right)    & \mbox{ otherwise }
\end{array}
\right.
\end{eqnarray*}

\noi
The proof is completed by placing the general expression for the
integrand's variance in (\ref{var1}) and (\ref{var2})
respectively. \hfill $\square$

\vspace{1.5cm}

\noi \textbf{Proof of Lemma \ref{var_cond_ind}}

\begin{eqnarray}
Var(\widehat{I}_{J})
&=& Var_{(\dn{u}, \dn{v})} \left\{ \frac{1}{R} \sum_{r=1}^{R} \left[
		  											\prod_{i=1}^{N}  \varphi_i \big(\textbf{\emph{u}}_i^{(r)},\textbf{\emph{v}}^{(r)}\big) \right] \right\} \nonumber \\
&=& \frac{1}{R} Var_{(\dn{u}, \dn{v})} \left[
		  											\prod_{i=1}^{N}  \varphi_i \big(\textbf{\emph{u}}_i,\textbf{\emph{v}}\big) \right]  \nonumber \\ 										
&=& \frac{1}{R} Var_{\dn{v}} \left\{ E_{\dn{u}|\dn{v}} \left[
		  											\prod_{i=1}^{N}  \varphi_i \big(\textbf{\emph{u}}_i,\textbf{\emph{v}}\big) \, \Big| \dn{v} \right]  \right\}
   +\frac{1}{R} E_{\dn{v}} \left\{ Var_{\dn{u}|\dn{v}} \left[
		  											\prod_{i=1}^{N}  \varphi_i \big(\textbf{\emph{u}}_i,\textbf{\emph{v}}\big) \, \Big| \dn{v} \right]  \right\}
\label{lemma33_eq1}											  																	
\end{eqnarray}

\noi
Due to conditional independence we have that
\begin{eqnarray}
E_{\dn{u}|\dn{v}} \left[ \prod_{i=1}^{N}  \varphi_i \big(\textbf{\emph{u}}_i,\textbf{\emph{v}}\big) \, \Big| \dn{v} \right]
&=&  \prod_{i=1}^{N}  E_{\dn{u}|\dn{v}} \left[  \varphi_i \big(\textbf{\emph{u}}_i,\textbf{\emph{v}}\big) \, \Big| \dn{v} \right]
 =   \prod_{i=1}^{N}  E \left(  \varphi_i  \big| \dn{v} \right).
\label{lemma33_eq2_cond_expectation}											  																	
\end{eqnarray}

\noi
Moreover, from (\ref{variGOOD2}) we have that
\begin{eqnarray}
Var_{\dn{u}|\dn{v}} \left[
		  											\prod_{i=1}^{N}  \varphi_i \big(\textbf{\emph{u}}_i,\textbf{\emph{v}}\big) \, \Big| \dn{v} \right]
&=&
\sum_{k=1}^{N} \sum_{ {\cal C} \in {{\cal N}\choose{k}} }
\Big[ \prod_{i \in {\cal C} } V\big( \varphi_i \big| \dn{v} \big) \prod_{j \in {\cal N} \setminus {\cal C} } E\big( \varphi_j \big| \dn{v} \big)^2 \Big]
\label{lemma33_eq3_var}											  																	
\end{eqnarray}

\noi
By substituting (\ref{lemma33_eq2_cond_expectation}) and (\ref{lemma33_eq3_var}) in (\ref{lemma33_eq1}),
we obtain the variance of the joint estimator of Lemma \ref{var_cond_ind}.

Similarly, for the marginal estimator we have
\begin{eqnarray}
Var \left( \widehat{\mathcal{I}}_M \right)
&=& Var_{(\dn{u}, \dn{v})} \left[\frac{1}{R_1} \sum_{r_1=1}^{R_1}\prod_{i=1}^N \overline{\varphi}_i^{(r_1)}  \right]
 =  \frac{1}{R_1}  Var_{(\dn{u}, \dn{v})} \Bigg[ \prod_{i=1}^N \overline{\varphi}_i  \Bigg]  \nonumber \\
&=&
\frac{1}{R_1}  Var_{\dn{v}} \left\{ E_{\dn{u}| \dn{v}} \Big[ \prod_{i=1}^N \overline{\varphi}_i  \, \Big| \dn{v} \Big] \right\}+
\frac{1}{R_1}  E_{\dn{v}} \left\{ Var_{\dn{u}| \dn{v}} \Big[ \prod_{i=1}^N \overline{\varphi}_i  \, \Big| \dn{v} \Big] \right\}
\label{lemma33_eq4_marg}											  																	
\end{eqnarray}

\noi
Due to conditional independence we have that
\begin{eqnarray}
 E_{\dn{u}| \dn{v}} \Big[ \prod_{i=1}^N \overline{\varphi}_i \, \Big| \dn{v}   \Big]
 &=&  \prod_{i=1}^{N}  E_{\dn{u}|\dn{v}} \left[  \overline{\varphi}_i \, \Big| \dn{v}  \right]
 =   \prod_{i=1}^{N}  E \left(  \varphi_i  \big| \dn{v} \right).
\label{lemma33_eq5}											  																	
\end{eqnarray}

Moreover, from Lemma \ref{prop:EstVars} we have that
\begin{eqnarray}
Var_{\dn{u}| \dn{v}} \Big[ \prod_{i=1}^N \overline{\varphi}_i  \, \Big| \dn{v} \Big]
&=&
\sum_{k=1}^{N} \left[ \frac{1}{R_2^k}  \sum_{ {\cal C} \in {{\cal N}\choose{k}} }
\prod_{i \in {\cal C} } V_i \prod_{j \in {\cal N} \setminus {\cal C} } E_j^2 \right],
\label{lemma33_eq6}
\end{eqnarray}

Substituting (\ref{lemma33_eq5}) and (\ref{lemma33_eq6}) in (\ref{lemma33_eq4_marg}) gives the expression of the
variance of the marginal estimator of Lemma \ref{var_cond_ind}.

\vspace{1.5cm}

\newpage
\noi \textbf{Proof of Lemma \ref{lemma:ISDI} }

The proof of Lemma \ref{lemma:ISDI}  can be  obtained by induction. The statement of the Lemma holds for $N=3$ with $\dn{Y}_3=(Y_1, Y_2, Y_3)$ since
\begin{eqnarray*}
&&Cov_{(3)}(\dn{Y}) + \sum_{k=1}^{1} \left[ \left( \prod^{3}_{i=4-k} \!\!\!\! E (Y_i) \right) Cov_{(3-k)}(\dn{Y}) \right]
= Cov_{(3)}(\dn{Y}) +   \left( \prod^{3}_{i=3 }  E (Y_i) \right) Cov_{(2)}(\dn{Y})  \\
&&\hspace{10em}= Cov( Y_1Y_2, Y_3 ) + E(Y_3) Cov(Y_1, Y_2) \\
&&\hspace{10em}= E( Y_1 Y_2 Y_3 ) - E(Y_1Y_2)E(Y_3) + E(Y_3) [ E(Y_1Y_2) - E(Y_1) E(Y_2)] \\
&&\hspace{10em}= TCI(\dn{Y}_3)~.
\end{eqnarray*}
which is true by the definition  of TCI (see equation \ref{covariation}) for vectors $\dn{Y}$ of length equal to three.

Let us now assume that (\ref{isdi1a}) it is true for any vector $\dn{Y}_N$ of length $N > 3$.
Then, for $\dn{Y}_{N+1}=( \dn{Y}_{N}, Y_{N+1} ) = ( Y_{1}, \dots,  Y_{N}, Y_{N+1} )$
the equation
\begin{equation}
\label{tci_N+1}
\mbox{TCI}(\dn{Y}_{N+1})=Cov_{(N+1)}(\dn{Y}) + \sum_{k=1}^{N-1} \left[ \left( \prod^{N+1 }_{i=N-k+2} \!\!\!\! E (Y_i) \right) Cov_{(N+1-k)}(\dn{Y}) \right] ,
\end{equation}
is also true since
\small
\begin{eqnarray*}
TCI( \dn{Y}_{N+1} ) \hspace{-0.7em}
&=& \hspace{-0.7em} E\left( \left[ \prod_{i=1}^N Y_i \right] \times Y_{N+1}\right) - \left[\prod_{i=1}^N E(Y_i)\right]  E(Y_{N+1}) \\
&=& Cov_{(N+1)}(\dn{Y}) + E\left( \prod_{i=1}^N Y_i \right) E(Y_{N+1})- \left[\prod_{i=1}^N E(Y_i)\right]  E(Y_{N+1})\\
&=& Cov_{(N+1)}(\dn{Y}) + TCI(\dn{Y}_N)  E(Y_{N+1})\\
&=& Cov_{(N+1)}(\dn{Y}) + \left\{ Cov_{(N)}(\dn{Y}) + \sum_{k=1}^{N-2} \left[ \left( \prod^{N }_{i=N-k+1} \!\!\!\! E (Y_i) \right) Cov_{(N-k)}(\dn{Y}) \right] \right\} E(Y_{N+1}) \\
&& \hspace{26em} \mbox{~~~ \it (from eq. \ref{isdi1a})} \\
&=& Cov_{(N+1)}(\dn{Y}) +   Cov_{(N)}(\dn{Y})E(Y_{N+1}) + \sum_{k=1}^{N-2} \left[ \left( \prod^{N+1 }_{i=N-k+1} \!\!\!\! E (Y_i) \right) Cov_{(N-k)}(\dn{Y}) \right]   \\
&=& Cov_{(N+1)}(\dn{Y}) +   Cov_{(N)}(\dn{Y})E(Y_{N+1}) + \sum_{k'=2}^{N-1} \left[ \left( \prod^{N+1 }_{i=N-k'+2} \!\!\!\! E (Y_i) \right) Cov_{(N-k'+1)}(\dn{Y}) \right]   \\
&& \hspace{25em} \mbox{\it( we set $k'=k+1$ )} \\
&=& Cov_{(N+1)}(\dn{Y}) +   \sum_{k'=1}^{N-1} \left[ \left( \prod^{N+1 }_{i=N-k'+2} \!\!\!\! E (Y_i) \right) Cov_{(N-k'+1)}(\dn{Y}) \right]  \\
&& \hfill \mbox{[{\it for $k=1$, the term  in the summation of (\ref{tci_N+1}) is equal to $Cov_{(N)}(\dn{Y})E(Y_{N+1})$}]}.
\end{eqnarray*}
\normalsize

\vspace{1.5cm}
\newpage
\noi \textbf{Proof of Lemma \ref{lemma:underestimaton}}

\begin{eqnarray}
Var \Big( \prod_{i=1}^N Y_i \Big)
&=&  E\left[ \prod_{i=1}^N  Y_i- E \Big(\prod_{i=1}^N  Y_i \Big)\right]^2 \nonumber \\
%&=&  E\left[ \prod_{i=1}^N  Y_i-TCI(\dn{Y})-\prod_{i=1}^N E( Y_i)\right]^2\nonumber \\
&=&   E\left[ \left (\prod_{i=1}^N Y_i-\prod_{i=1}^N E (Y_i) \right)-TCI(\dn{Y})\right]^2\nonumber \\
&=&   E\left[ \prod_{i=1}^N Y_i-\prod_{i=1}^N E (Y_i)\right]^2 +TCI(\dn{Y})^2 - 2 \, E\left\{TCI(\dn{Y}) \Big[\prod_{i=1}^N Y_i-\prod_{i=1}^N E (Y_i)\Big]\right\}  \nonumber \\
&=&   E\left[ \prod_{i=1}^N Y_i-\prod_{i=1}^N E (Y_i)\right]^2
 =   Var\left( \prod_{i=1}^N Y_i \Big| Independence \right) -TCI(\dn{Y})^2.\nonumber
\end{eqnarray}
since
$E\left\{TCI(\dn{Y}) \Big[\prod_{i=1}^N Y_i-\prod_{i=1}^N E (Y_i)\Big]\right\}
=TCI(\dn{Y}) E\Big[\prod_{i=1}^N Y_i-\prod_{i=1}^N E (Y_i)\Big] = 0$.
\hfill $\square$

\newpage

\bibliography{phdlit}

\begin{thebibliography}{}

\bibitem[Aguilar and West, 2000]{agwest00}
Aguilar, O. and West, M. (2000).
\newblock Bayesian {D}ynamic {F}actor {M}odels and portfolio allocation.
\newblock {\em Journal of Business and Economic Statistics}, 18:338--357.

\bibitem[Baker, 1998]{Baker98}
Baker, F. (1998).
\newblock An investigation of the item parameter recovery characteristics of a
  {G}ibbs sampling procedure.
\newblock {\em Applied Psychological Measurement}, 22:153--169.

\bibitem[Bartholomew et~al., 2011]{bart2011}
Bartholomew, D., Knott, M., and Moustaki, I. (2011).
\newblock {\em Latent variable models and factor analysis: a unified approach}.
\newblock Wiley Series on Probability and Statistics. John Wiley and Sons,
  London, {UK}, 3rd edition.

\bibitem[Bock and Aitkin, 1981]{BockAit81}
Bock, R. and Aitkin, M. (1981).
\newblock Marginal maximum likelihood estimation of item parameters:
  {A}pplication of an {EM} algorithm.
\newblock {\em Psychometrika}, 46:443--459.

\bibitem[Bock and Lieberman, 1970]{BockLie70}
Bock, R.~D. and Lieberman, M. (1970).
\newblock Fitting a response model for n dichotomously scored items.
\newblock {\em Psychometrika}, 35:179--197.

\bibitem[Bratley et~al., 1987]{bratetal:1987}
Bratley, P., Fox, B.~L., and Schrage, L. (1987).
\newblock {\em A guide to simulation}.
\newblock Springer, second edition.

\bibitem[Carlin and Louis, 2000]{CarLouis00}
Carlin, B.~P. and Louis, T.~A. (2000).
\newblock {\em Bayes and {E}mpirical {B}ayes methods for data analysis}.
\newblock Chapman \& Hall/CRC, second edition.

\bibitem[DiCiccio et~al., 1997]{diccetal97}
DiCiccio, T.~J., Kass, R.~E., Raftery, A., and Wasserman, L. (1997).
\newblock Computing {B}ayes {F}actors by combining simulation and asymptotic
  approximations.
\newblock {\em Journal of the American Statistical Association},
  92(439):903--915.

\bibitem[Fouskakis et~al., 2009]{fousk:09}
Fouskakis, D., Ntzoufras, I., and Draper, D. (2009).
\newblock Bayesian variable selection using cost-adjusted {BIC}, with
  application to cost-effective measurement of quality of health care.
\newblock {\em Annals of Applied Statistics}, 3:663--690.

\bibitem[Gelfand and Dey, 1994]{GelDey94}
Gelfand, A.~E. and Dey, D.~K. (1994).
\newblock Bayesian {M}odel {C}hoice: Asymptotics and exact calculations.
\newblock {\em Journal of the Royal Statistical Society. Series B
  (Methodological)}, 56(3):501--514.

\bibitem[Gelman and Meng, 1998]{ge:me98}
Gelman, A. and Meng, X.-L. (1998).
\newblock Simulating normalizing constants: From {I}mportance sampling to
  {B}ridge sampling to {P}ath sampling.
\newblock {\em Statistical Science}, 13(2):163--185.

\bibitem[Geweke and Zhou, 1996]{geZou:96}
Geweke, J. and Zhou, G. (1996).
\newblock Measuring the pricing error of the {A}rbitrage {P}ricing {T}heory.
\newblock {\em Review of Financial Studies}, 9:557--587.

\bibitem[Gifford and Swaminathan, 1990]{GiffSwa90}
Gifford, J.~A. and Swaminathan, H. (1990).
\newblock Bias and the effect of priors in {B}ayesian estimation of parameters
  of {I}tem {R}esponse {M}odels.
\newblock {\em Applied Psychological Measurement}, 14:33--43.

\bibitem[Goodman, 1962]{good62}
Goodman, L.~A. (1962).
\newblock The variance of the product of {K} random variables.
\newblock {\em Journal of the American Statistical Association}, 57:54--60.

\bibitem[Huber et~al., 2004]{huber.ea:04}
Huber, P., Ronchetti, E., and Victoria-Feser, M.-P. (2004).
\newblock Estimation of generalized linear latent variable models.
\newblock {\em Journal of the Royal Statistical Society, Series B},
  66:893--908.

\bibitem[Jones et~al., 2006]{Joetal06}
Jones, G., Haran, M., Caffo, B., and Neath, R. (2006).
\newblock Fixed-width output analysis for {M}arkov {C}hain {M}onte {C}arlo.
\newblock {\em Journal of the American Statistical Association},
  101:1537--1547.

\bibitem[Kang and Cohen, 2007]{KanCo07}
Kang, T. and Cohen, A.~S. (2007).
\newblock Irt model selection methods for dichotomous items.
\newblock {\em Applied Psychological Measurement}, 31(4):331–358.

\bibitem[Kass and Raftery, 1995]{kas:raf95}
Kass, R. and Raftery, A. (1995).
\newblock Bayes factors.
\newblock {\em Journal of the American Statistical Association}, 90:773--795.

\bibitem[Kim et~al., 1994]{Kimetal94}
Kim, S.-H., Cohen, A.~S., Baker, F.~B., Subkoviak, M.~J., and Leonard, T.
  (1994).
\newblock An investigation of hierarchical bayes procedures in item response
  theory.
\newblock {\em Psychometrika}, 59(3):405--421.

\bibitem[Koehler et~al., 2009]{koeh09}
Koehler, E., Brown, E., and Haneuse, S. J.-P.~A. (2009).
\newblock On the assessment of {M}onte {C}arlo error in simulation-based
  statistical analyses.
\newblock {\em The American Statistician}, 63(2):155--162.

\bibitem[Lewis and Raftery, 1997]{le:raf97}
Lewis, S. and Raftery, A. (1997).
\newblock Estimating {B}ayes factors via posterior simulation with the
  {L}aplace {M}etropolis estimator.
\newblock {\em Journal of the American Statistical Association}, 92:648--655.

\bibitem[Lopes and West, 2004]{lowest:04}
Lopes, H.~F. and West, M. (2004).
\newblock Bayesian model assessment in factor analysis.
\newblock {\em Statistica Sinica}, 14:41–67.

\bibitem[Lord, 1980]{Lord1980}
Lord, F.~M. (1980).
\newblock {\em {A}pplications of {I}tem {R}esponse {T}heory to practical
  testing problems}.
\newblock Erlbaum Associates, Hillsdale, NJ.

\bibitem[Lord and Novick, 1968]{LorNov68}
Lord, F.~M. and Novick, M.~R. (1968).
\newblock {\em Statistical theories of mental test scores}.
\newblock Addison-Wesley, Oxford, UK.

\bibitem[Meng and Schilling, 2002]{me:sch02}
Meng, X.-L. and Schilling, S. (2002).
\newblock Warp {B}ridge {S}ampling.
\newblock {\em Journal of Computational and Graphical Statistics},
  11(3):552--586.

\bibitem[Meng and Wong, 1996]{me:wo96}
Meng, X.-L. and Wong, W.-H. (1996).
\newblock Simulating ratios of normalizing constants via a simple identity: A
  theoretical exploration.
\newblock {\em Statistica Sinica}, 6:831--860.

\bibitem[Mislevy, 1986]{Misl86}
Mislevy, R. (1986).
\newblock Bayes modal estimation in {I}tem {R}esponse {M}odels.
\newblock {\em Psychometrika}, 51:177--195.

\bibitem[Moustaki and Knott, 2000]{mou:kn00}
Moustaki, I. and Knott, M. (2000).
\newblock Generalized {L}atent {T}rait {M}odels.
\newblock {\em Psychometrika}, 65:391--411.

\bibitem[Ntzoufras et~al., 2003]{ntzetal:03}
Ntzoufras, I., Dellaportas, P., and Forster, J. (2003).
\newblock Bayesian variable and link determination for {G}eneralised {L}inear
  {M}odels.
\newblock {\em Journal of Statistical Planning and Inference},
  111(1-2):165--180.

\bibitem[Patz and Junker, 1999]{PatzJunk99}
Patz, R. and Junker, B. (1999).
\newblock A straightforward approach to {M}arkov {C}hain {M}onte {C}arlo
  methods for {I}tem {R}esponse {M}odels.
\newblock {\em Journal of Educational and Behavioral Statistics}, 24:146--178.

\bibitem[Rabe-Hesketh et~al., 2005]{rabe.skrondal.pickles:05}
Rabe-Hesketh, S., Skrondal, A., and Pickles, A. (2005).
\newblock Maximum likelihood estimation of limited and discrete dependent
  variable models with nested random effects.
\newblock {\em Journal of Econometrics}, 128:301--323.

\bibitem[Schilling and Bock, 2005]{schilling.Bock:05}
Schilling, S. and Bock, R. (2005).
\newblock High-dimensional maximum marginal likelihood item factor analysis by
  adaptive quadrature.
\newblock {\em Psychometrika}, 70:533--555.

\bibitem[Schmeiser, 1982]{Schm:82}
Schmeiser, B.~W. (1982).
\newblock Batch size effects in the analysis of simulation output.
\newblock {\em Operations Research}, 30:556--568.

\end{thebibliography}
\end{document}